\documentclass[12pt]{article}
\usepackage[dvips]{color}
\usepackage{epsfig}
\usepackage{amsmath}
\usepackage{subfigure}
\usepackage{graphicx}

\begin{document}
\title{\bf Higher order corrections of the extended Chaplygin gas cosmology with varying $G$ and $\Lambda$}
\author{{\small {E.O. Kahya$^{a}$\thanks{Email: eokahya@itu.edu.tr}, M. Khurshudyan$^{b}$\thanks{Email: martiros.khurshudyan@mpikg.mpg.de}, B. Pourhassan$^{c}$\thanks{Email: bpourhassan@yahoo.com}, R. Myrzakulov$^{d}$\thanks{Email: rmyrzakulov@gmail.com}, A. Pasqua$^{e}$\thanks{Email: toto.pasqua@gmail.com}}}\\
$^{a}${\small {\em Physics Department, Istanbul Technical University, Istanbul, Turkey}}\\
$^{b}${\small {\em Max Planck Institute of Colloids and Interfaces,}}\\
{\small {\em Potsdam-Golm Science Park Am Mühlenberg 1 OT Golm 14476 Potsdam}}\\
$^{c}${\small {\em School of Physics, Damghan University, Damghan, Iran}}\\
$^{d}${\small {\em Eurasian International Center for Theoretical Physics, Eurasian National University,}}\\
{\small {\em Astana 010008, Kazakhstan}}\\
$^{e}${\small {\em Department of Physics, University of Trieste, Via Valerio, 2 34127 Trieste, Italy}}}  \maketitle
\begin{abstract}
In this paper, we study two different models of dark energy based on Chaplygin gas equation of state. The first model is the variable modified Chaplygin gas while the second one is the extended Chaplygin gas. Both models are considered in the framework of higher order $f(R)$ modified gravity. We also consider the case of time varying gravitational constant $G$ and $\Lambda$ for both models. We investigate some cosmological parameters such as the Hubble, the deceleration and the equation of state parameters. Then we showed that the model that we considered, extended Chaplygin gas with time-dependent $G$ and $\Lambda$,
is consistent with the observational data. Finally we conclude with the discussion of cosmological perturbations of our model.
\end{abstract}

\section{Introduction}
Cosmological and astrophysical data obtained thanks to the Supernovae Ia  (SNeIa), the Cosmic Microwave Background (CMB) radiation anisotropies, the Large Scale Structure (LSS) and X-ray experiments provide strong evidences supporting a phase of accelerated expansion of the present day universe \cite{1,1a,1b,1c,1d,cmb1,cmb2,cmb3,planck,sds1,sds2,sds4,xray}.\\
In order to find a suitable model for our universe, some possible reasons for  this observed accelerated expansion have been investigated. Three main classes of models are usually proposed to describe this phenomenon: a) a  cosmological constant $\Lambda$; b) Dark Energy (DE) models; c) modified theories of gravity.\\
The cosmological constant $\Lambda$, which solves the equation of state (EoS) parameter $\omega = -1$,  represents the simplest candidate proposed to explain the accelerated expansion of the universe. However, $\Lambda$ is well known to be related to the fine-tuning and the cosmic coincidence problems \cite{copeland-2006,P5}. According to the first, the vacuum energy density is about $10^{123}$ times smaller than what is  observed. Instead, according to the cosmic coincidence problem, the vacuum energy and Dark Matter (DM) are nearly equal nowadays, although they have evolved independently and from different mass scales (which is a particular fortuity if no interaction exist between them). Many attempts have been made with the aim to propose a solution to the coincidence problem \cite{delcampo,delcampoc,delcampoe,delcampol}.\\
DE may explain the origin of the accelerated expansion of universe \cite{P1, P2}. There are several models to describe the dark energy such as the models based on quintessence \cite{P3,P4,41}. K-essence \cite{P6} and tachyonic models \cite{P7} are also two among many other ways to describe dark energy.\\
A successful model to describe DE is based on Chaplygin gas (CG) equation of state \cite{P7,P8} and yields the Generalized Chaplygin Gas (GCG) model \cite{P9,P10}. It was initially emerged in cosmology from string theory point of view \cite{john1, john2, P1 0003288}.
One can indeed unify dark matter and dark energy using this model. It is also possible to study effect of viscosity in GCG \cite{P15,P16,P17}.
However, observational data ruled out such a proposal, and the modified Chaplygin gas (MCG) model introduced \cite{P21}. Recently, viscous MCG is also suggested and studied (\cite{P22,P23}. A further extension of CG model is called modified cosmic Chaplygin gas (MCCG) which was recently proposed \cite{P24,P25,P26,P27}.  Moreover, various Chaplygin gas models were studied from the holography point of view \cite{P28,P29,P30}.\\
The MCG equation of state has two parts, the first term gives an ordinary fluid obeying a linear barotropic EoS while the second term relates the pressure to some power of the inverse of energy density. Therefore, we are dealing with a two-fluid model. However, it is possible to consider barotropic fluid with quadratic EoS or even with higher orders EoS \cite{P31,P32}. Therefore, it is interesting to extend MCG EoS which recovers at least barotropic fluid with quadratic EoS. We called the new version as the extended Chaplygin gas (ECG) model \cite{RP2014,ASS2014}. In order to get better agreement with observational data one can develop interesting models by varying the constants in EoS parameter.\\
The cosmic acceleration has been also accurately studied by imposing the concept of modification of gravity  \cite{nojod}. This new model of gravity (predicted by string/M theory) gives a very natural gravitational alternative for exotic matter. The explanation of the phantom, non-phantom and quintom phases of the universe can be well described using modified gravity without the necessity of the introduction of a negative kinetic term in DE models.
The cosmic  acceleration is evinced by the straightforward fact that terms like $1/R$ might become fundamental at small curvature. Furthermore, modified gravity models provide a natural way to join the early-time inflation and late-time acceleration. Such theories are also prime candidates for the explanation of the DE and DM, including for instance the anomalous galaxies rotation curves. The effective DE dominance may be assisted by the modification of gravity. Hence, the coincidence problem is solved there simply by the fact that the universe expands.\\
Modified gravity is also expected to be useful in high energy physics, explaining the hierarchy problem or unification of other forces with gravity \cite{nojod}. Some of the most famous and known models of modified gravity are represented by braneworld models, $f(T)$ gravity (where $T$ indicates the torsion scalar),
$f(R)$ gravity (where $R$ indicates the Ricci scalar curvature), $f(G)$ gravity where
$$G=R^2-4R_{\mu \nu}R^{\mu \nu} + R_{\mu \nu \lambda \sigma}R^{\mu \nu \lambda \sigma}$$
is the Gauss-Bonnet invariant, with $R_{\mu \nu}$ representing the Ricci curvature tensor and $R_{\mu \nu \lambda \sigma}$ representing the Riemann curvature tensor, $f(R,T)$ gravity, DGP models, DBI models and Ho${\check{\rm r}}$ava-Lifshitz gravity \cite{15g,15h,15i,mio,bra1,bra2,mioft1,ft4,ft8,fr2,miofr,fr9,fr8,miofg1,frt2,miodbi,miohl,miobd2}.\\
In order to obtain a comprehensive model, we also add two modifications to the ordinary model. First, we consider a fluid which governs the background dynamics of the universe in a higher derivative theory of gravity. Second, we consider time varying $G$ and $\Lambda$. As we know, the Einstein equations of general relativity do not permit any variations in the gravitational constant $G$ and cosmological constant $\Lambda$. Since the Einstein tensor has zero divergence and energy conservation law is also zero, some modifications of Einstein equations are required. A similar study has been recently performed for another fluid model instead of Chaplygin gas \cite{P33}. There are also several works on cosmological models with varying $G$ and $\Lambda$  \cite{P34,P35}.\\
Therefore, in this paper, we study two different models of Chaplygin gas models in higher order gravity with varying $G$ and $\Lambda$.\\
This paper is organized as follows. In section 2, we introduce our models. In section 3, we study special cases corresponding to constant $G$ and $\Lambda$.
In section 4, we investigate the Hubble, the deceleration and the EoS parameters for the two models that were introduced. In Section 5, we consider the statefinder diagnostics for both models. In section 6, we make a perturbation analysis for Chaplygin gas. Finally, in section 7, we write the conclusions of this paper.
\section{The models}
We consider two different models for a fluid which governs the background dynamics of the universe in a higher derivative theory of gravity in the presence of time varying $G$ and $\Lambda$. Within modified theories of gravity, we hope to solve the problems of the dark energy which are originated from general relativity.\\
A gravitational action with higher order term in the Ricci scalar curvature $R$ containing a variable gravitational constant $G(t)$ is given by:
\begin{equation}\label{1eq:Lag}
I = - \int {d^{4}x \sqrt{-g}\left [ \frac{1}{ 16 \pi G} f(R) +L_{m} \right ] },
\end{equation}
where $f(R)$ is a function of $R$ and its higher power (including a variable $\Lambda$), $g$ is the determinant of the four dimensional tensor metric $g^{\mu \nu}$ and $L_{m}$ represents the matter Lagrangian. Considering the second order gravity, we can take into account the following expression of $f(R)$:
\begin{equation}\label{2}
f(R)=R + \alpha R^{2}-2\Lambda,
\end{equation}
where $\alpha$ is a constant parameter.\\
By using the following flat FRW metric:
\begin{equation}\label{3eq:s2}
ds^2=-dt^2+a(t)^2\left(dr^{2}+r^{2}d\Omega^{2}\right),
\end{equation}
where $d\Omega^{2}=d\theta^{2}+\sin^{2}\theta d\phi^{2}$ represents the angular part of the metric and $a(t)$
represents the scale factor (which gives information about the expansion of the universe), we get the following Friedmann equations \cite{Paul}:
\begin{equation}\label{4eq:Fridmmanvlambda}
H^{2}-6\alpha(2H\ddot{H}-\dot{H}^{2}+6\dot{H}H^{2})=\frac{8\pi G}{3}\rho+\frac{\Lambda}{3},
\end{equation}
\begin{equation}\label{4eq:Fridmmanvlambda2}
\dot{H}+H^{2}-6\alpha(2H\ddot{H}-\dot{H}^{2}+6\dot{H}H^{2})=-\frac{4\pi G}{3}(3P+\rho)+\frac{\Lambda}{3}
\end{equation}
where an overdot and two overdots indicate, respectively, the first and the second derivative with respect to the cosmic time $t$.\\
The energy conservation equation is given by the following relation:
\begin{equation}\label{5eq:conservation}
\dot{\rho}+3H(\rho+P)=-\left( \frac{\dot{G}}{G}\rho +\frac{\dot{\Lambda}}{8\pi G}\right ),
\end{equation}
where $\rho$ and $P$ are the energy density and the pressure of the perfect fluid, respectively.
Modified theories of gravity (like $f(R)$ theories) give the opportunity to find a natural representation and
introduction of the dark energy into theory. Therefore, the type of the dark energy and dynamics of the universe depends on
the form of $f(R)$ which will be considered. The type of the work which we would like to consider in this paper assume
the existence of an effective fluid controlling the dynamics of the universe composed non-interacting dark energy
(emerging from $f(R)$) and a fluid emerging from our assumptions.\\
Assuming there is not interaction, for matter we have the following continuity equation:
\begin{equation}\label{6eq:drho}
\dot{\rho}+3H(\rho+P)=0.
\end{equation}
Therefore, comparing Eqs. (\ref{5eq:conservation}) and (\ref{6eq:drho}), we can easily derive the following relation for the dynamics of $G$:
\begin{equation}\label{7}
\dot{G} +\frac{\dot{\Lambda}}{8\pi \rho}=0.
\end{equation}
The energy density $\rho$ can be assumed as originated by some kinds of Chaplygin gas. In particular, we have that the MCG is described by the following EoS:
\begin{equation}\label{8}
p=A\rho-\frac{B}{\rho^{n}},
\end{equation}
where $A$ and $B$ are two arbitrary constant parameters which may fitted using observational data. The special case corresponding to $A = 0$ yields to the GCG EoS, while the special case corresponding to $A = 0$ and $n = 1$ recovers the pure Chaplygin gas EoS. Moreover, the limiting case corresponding to $n=0.5$ has been studied in \cite{P84}. Then, authors of the Ref. \cite{P85} concluded that the best fitted parameters are $A = -0.085$ and $\alpha = 1.724$, while Constitution + CMB + BAO data suggests $A = 0.061 \pm 0.079$, $n = 0.053 \pm 0.089$, and Union + CMB + BAO results suggests $A = 0.110 \pm 0.097$, $n = 0.089 \pm 0.099$ \cite{P86}. Other observational constraints on MCG model using Markov Chain Monte Carlo suggests $A = 0.00189^{+0.00583}_{-0.00756}$, $\alpha=0.1079^{+0.3397}_{-0.2539}$ at $1\sigma$ level and $A = 0.00189^{+0.00660}_{-0.00915}$, $n=0.1079^{+0.4678}_{-0.2911}$ at $2\sigma$ level \cite{P87}. There are also other constraints for example those reported by the Refs. \cite{P88,P89,P90,P91}.\\
It is possible to consider $B$ as a variable (depending on the time $t$) instead of a constant. So, a time-varying MCG can be described by the following EoS:
\begin{equation}\label{9}
P=A\rho-\frac{B(t)}{\rho^{n}},
\end{equation}
where
\begin{equation}\label{10}
B(t)=\omega(t)a(t)^{-3(1-\omega(t))(1-n)},
\end{equation}
with  $\omega(t)$ given by \cite{0801.4529}:
\begin{equation}\label{11}
\omega(t)=\omega_{0}+\omega_{1}t\frac{\dot{H}}{H}.
\end{equation}
where $\omega_0$ and $\omega_1$ are two constant parameters.\\
In the second model we consider ECG with the following EoS:
\begin{equation}\label{12}
P=\sum_{k=1}^{m}{A_{k}\rho^{k}}-\frac{B}{\rho^{n}},
\end{equation}
$B$ and $n$ are arbitrary constants, and $A_{k}=1/k$ assumed in this paper. The ECG EoS reduces to the MCG EoS in the limiting case of $k=1$, and can recover barotropic fluid with quadratic EoS by setting $k=2$. Moreover, higher values of $k$ may recover higher order barotropic fluid which is indeed our motivation to consider the ECG.

\section{Numerical results with constant $G$ and $\Lambda$}
We start the analysis of the models considered in this paper with the case of constant $\Lambda$ and $G$ (we will also consider units of $8\pi G=c=1$). Therefore, we have the following two equations, which will describe the dynamics of the universe, in this case the dynamics of the Chaplygin gases described by the following equations:
\begin{equation}\label{13eq:Fridmmanvlambdaa}
H^{2}-6\alpha(2H\ddot{H}-\dot{H}^{2}+6\dot{H}H^{2})=\frac{\rho}{3}+\frac{\Lambda}{3},
\end{equation}
and
\begin{equation}\label{14eq:constlambda}
\dot{\rho}+3H(\rho+P)=0.
\end{equation}

\begin{figure}[h!]
 \begin{center}$
 \begin{array}{cccc}
\includegraphics[width=60 mm]{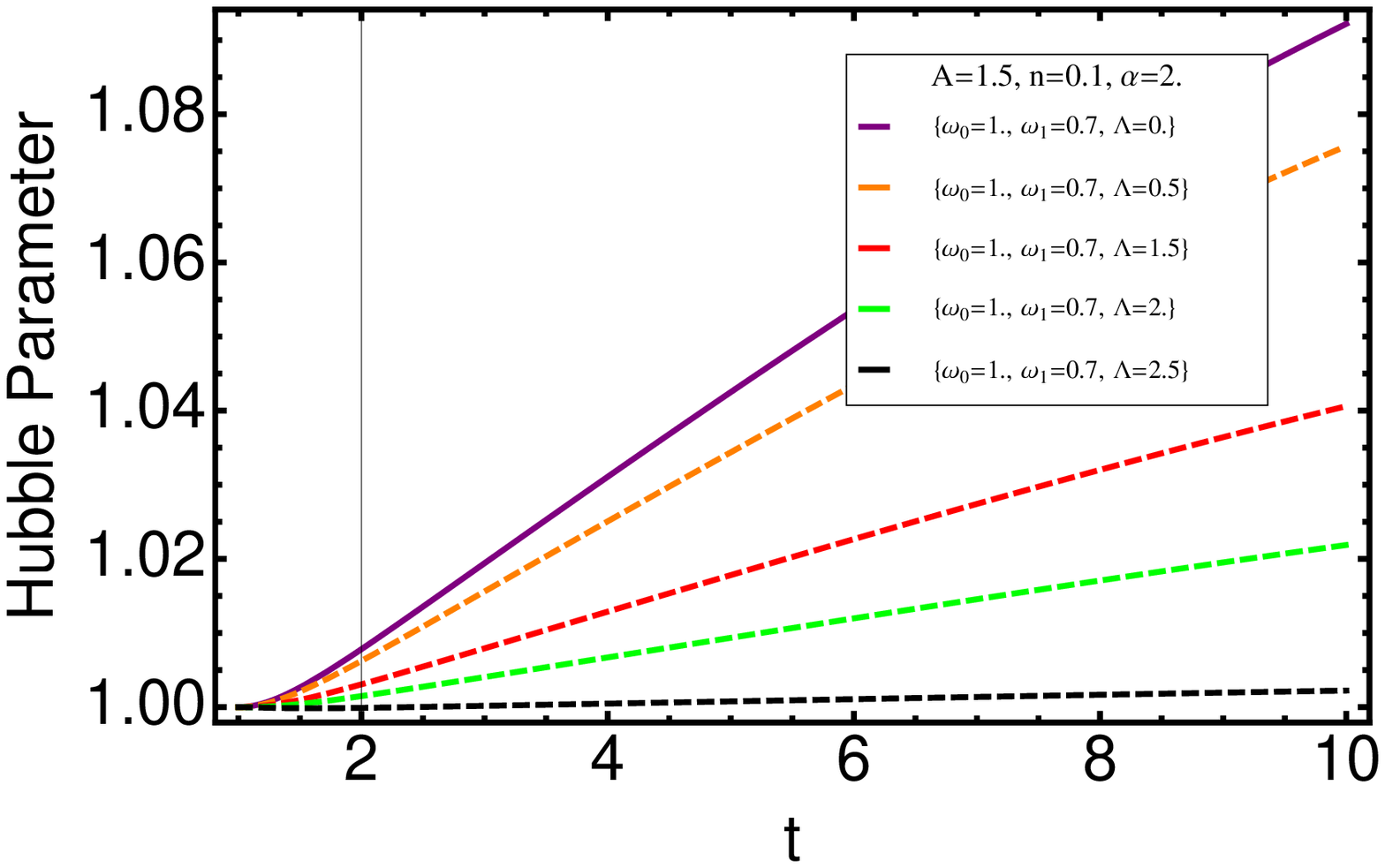} &
\includegraphics[width=60 mm]{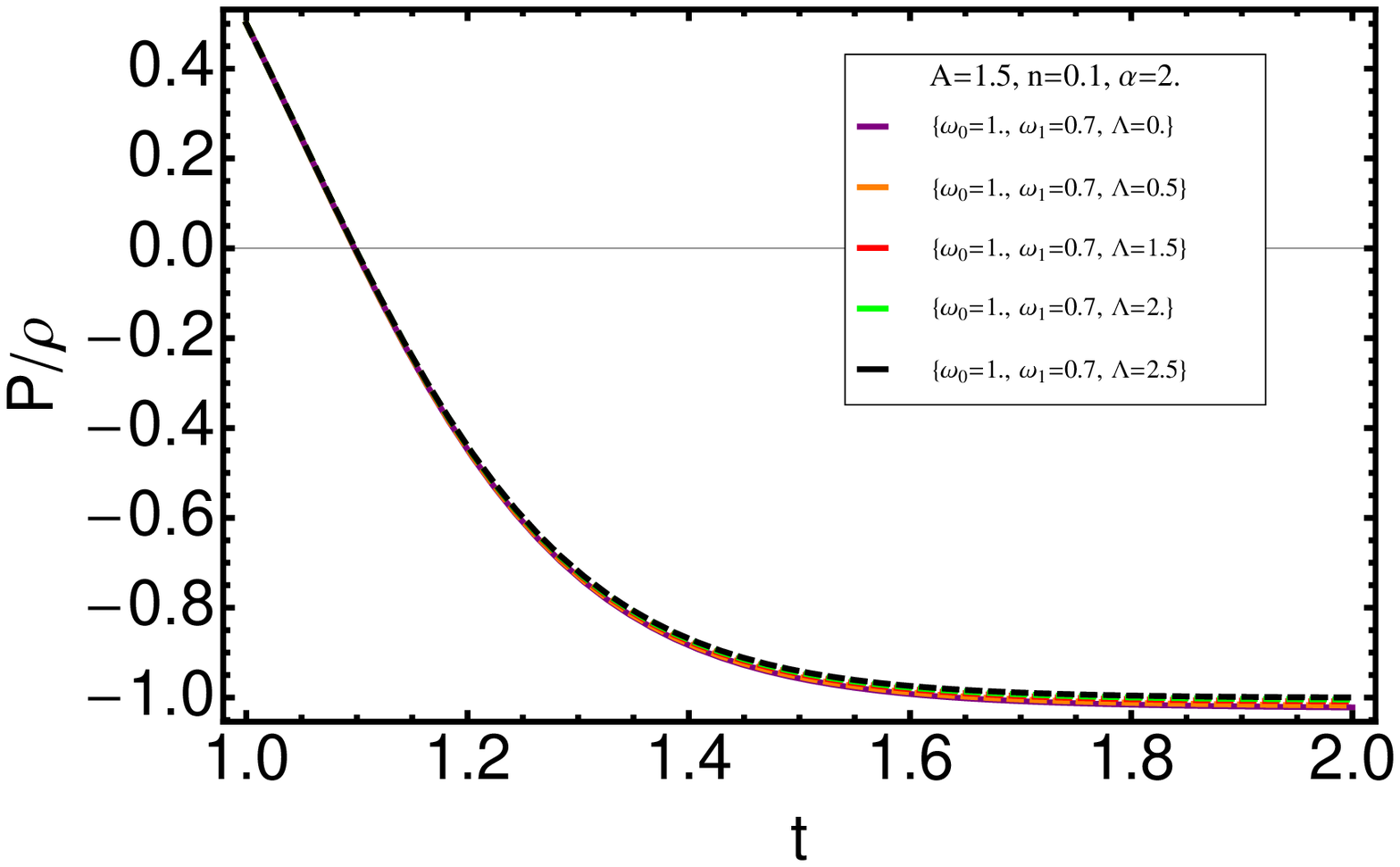}
 \end{array}$
 \end{center}
\caption{Behavior of $H$ and $\omega$ against $t$.}
 \label{fig:1}
\end{figure}

\begin{figure}[h!]
 \begin{center}$
 \begin{array}{cccc}
\includegraphics[width=60 mm]{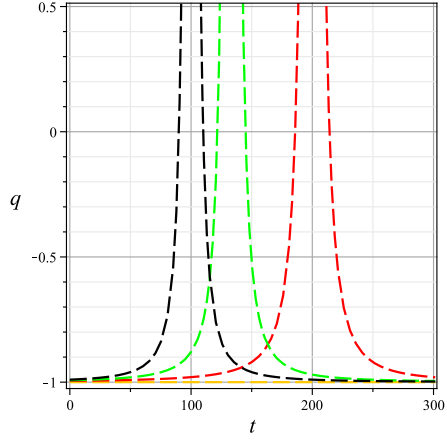}
 \end{array}$
 \end{center}
\caption{Behavior of $q$ against $t$. $\Lambda=2.5$ (orange), $\Lambda=3.5$ (red), $\Lambda=4$ (green), $\Lambda=5$ (black).}
 \label{fig:2}
\end{figure}

In the first model, given by the EoS (\ref{9}), we obtain the behavior of the Hubble parameter and the EoS parameter $\omega=P/\rho$ by using a numerical analysis which is represented in Fig. \ref{fig:1}. The left plot of Fig. \ref{fig:1} shows a typical time evolution of the Hubble parameter. It is found that  the value of $H$ decreases with the increasing  of the value of $\Lambda$. For $\Lambda<2.5$, the Hubble parameter is an increasing function of time. $\Lambda=2.5$ yields to the constant $H$ while for $\Lambda>2.5$  the Hubble parameter is a decreasing function of the time.\\
Since one would expect the Hubble parameter to decrease with time and become constant at the present epoch,
this model is not in good agreement with observational data.
However, the evolution of the EoS parameter (right plot of Fig. \ref{fig:1}) agrees with the $\Lambda$CDM model where it is
expected to behave as $\omega\rightarrow-1$.\\
Using Fig. \ref{fig:1}, we can obtain the following fit of the function of the Hubble parameter:
\begin{equation}\label{15}
H=H_{0}+C(2.5-\Lambda)t,
\end{equation}
where $H_{0}$ is the current value of the Hubble parameter and $C$ is a constant. Using the expression of $H$ obtained in Eq. (\ref{15}), we can now investigate the deceleration parameter $q$ via the following relation:
\begin{equation}\label{16}
q=-1-\frac{\dot{H}}{H^{2}} = - 1 -\frac{C(2.5-\Lambda)}{\left[ H_{0}+C(2.5-\Lambda)t   \right]^2}.
\end{equation}
Numerical analysis yields to Fig. \ref{fig:2} which shows that, for appropriate values of parameters involved, we obtain $q=-1$ which is in agreement with the $\Lambda$CDM model. In Fig. \ref{fig:2}, we can also see deceleration to acceleration and acceleration to deceleration phase transitions. Curves of this figure drawn for $\Lambda\geq2.5$, however in the cases of $\Lambda\leq2.5$ we find $q\leq-1$.

\begin{figure}[h!]
 \begin{center}$
 \begin{array}{cccc}
\includegraphics[width=55 mm]{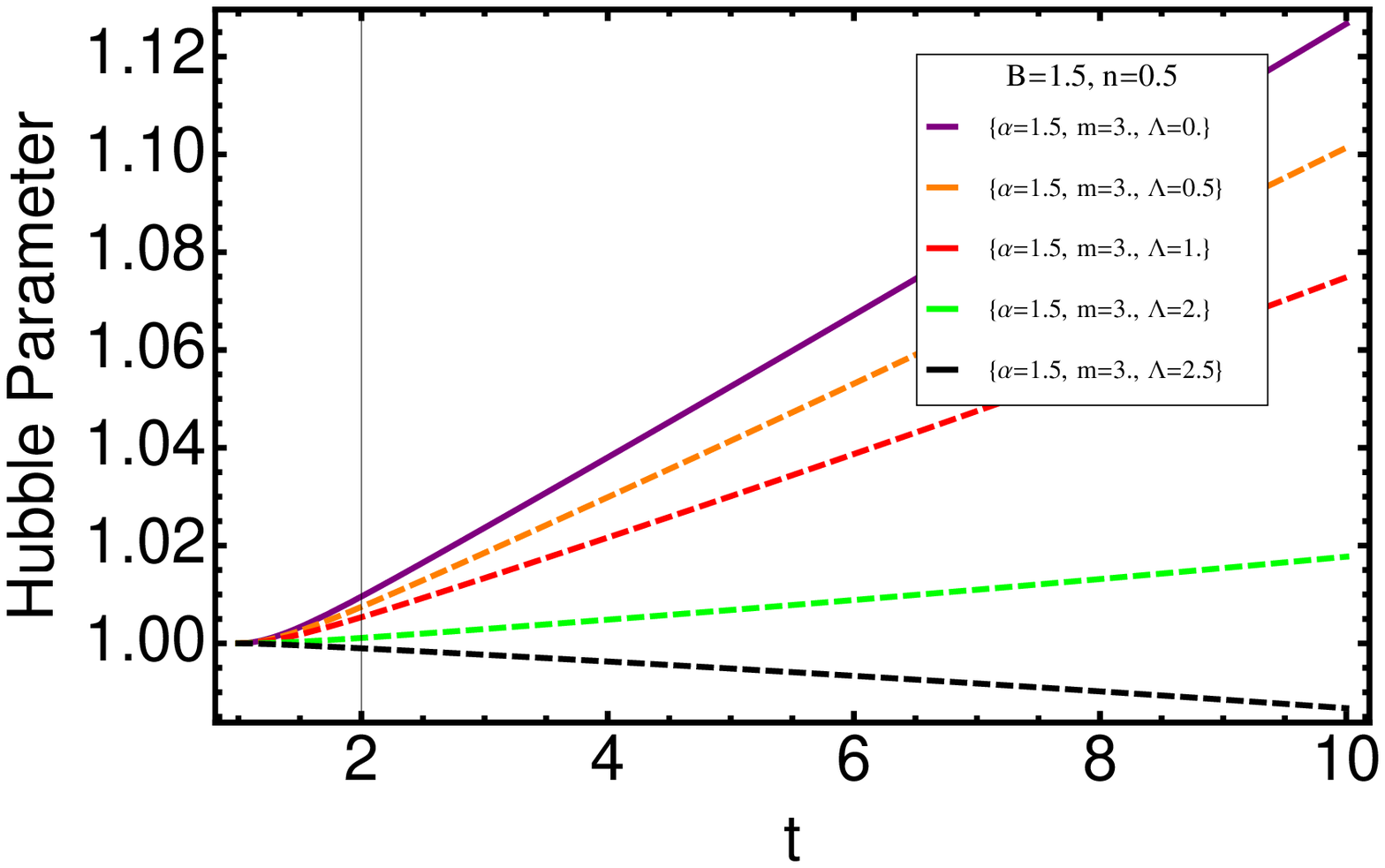} &
\includegraphics[width=55 mm]{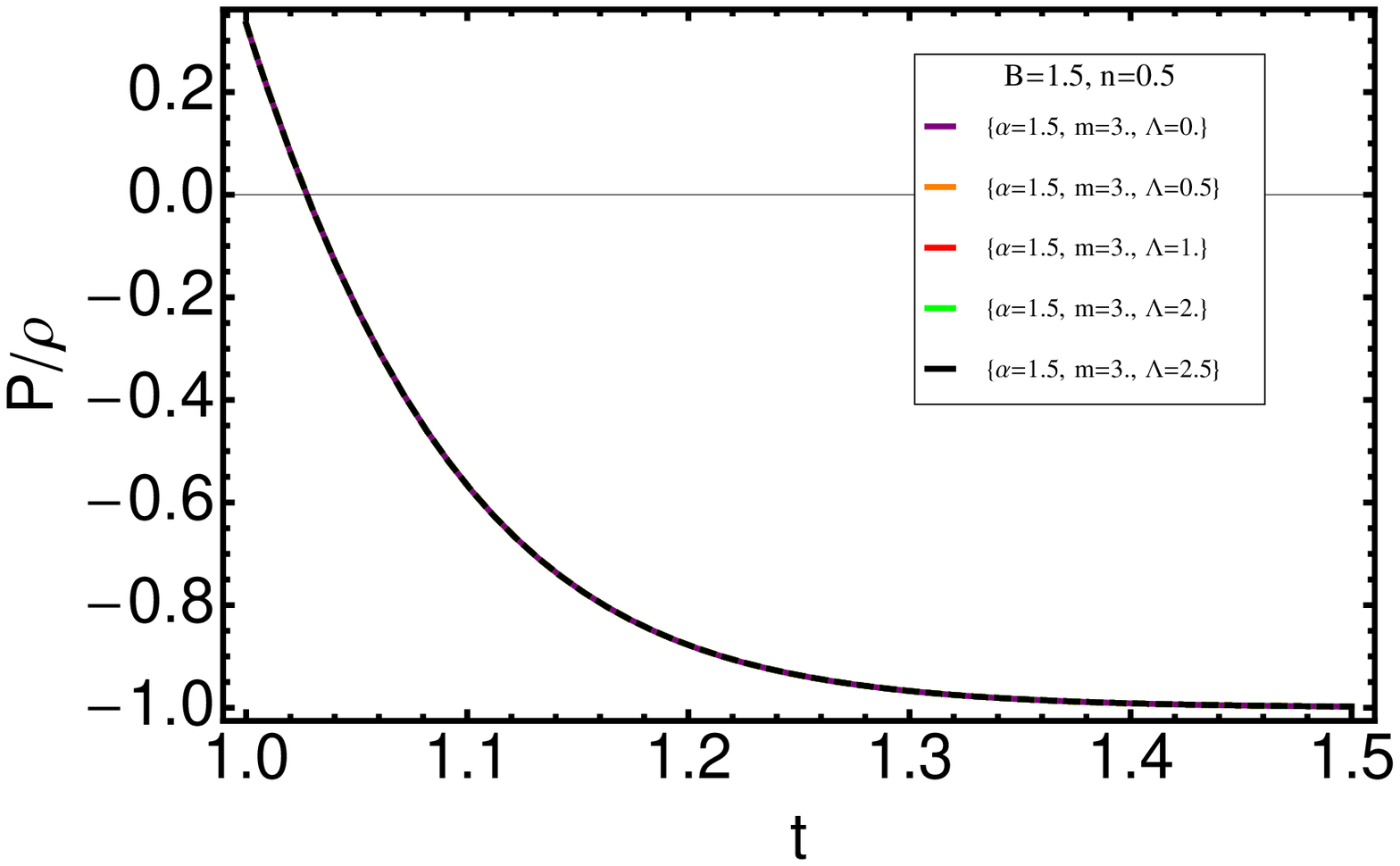}&
\includegraphics[width=55 mm]{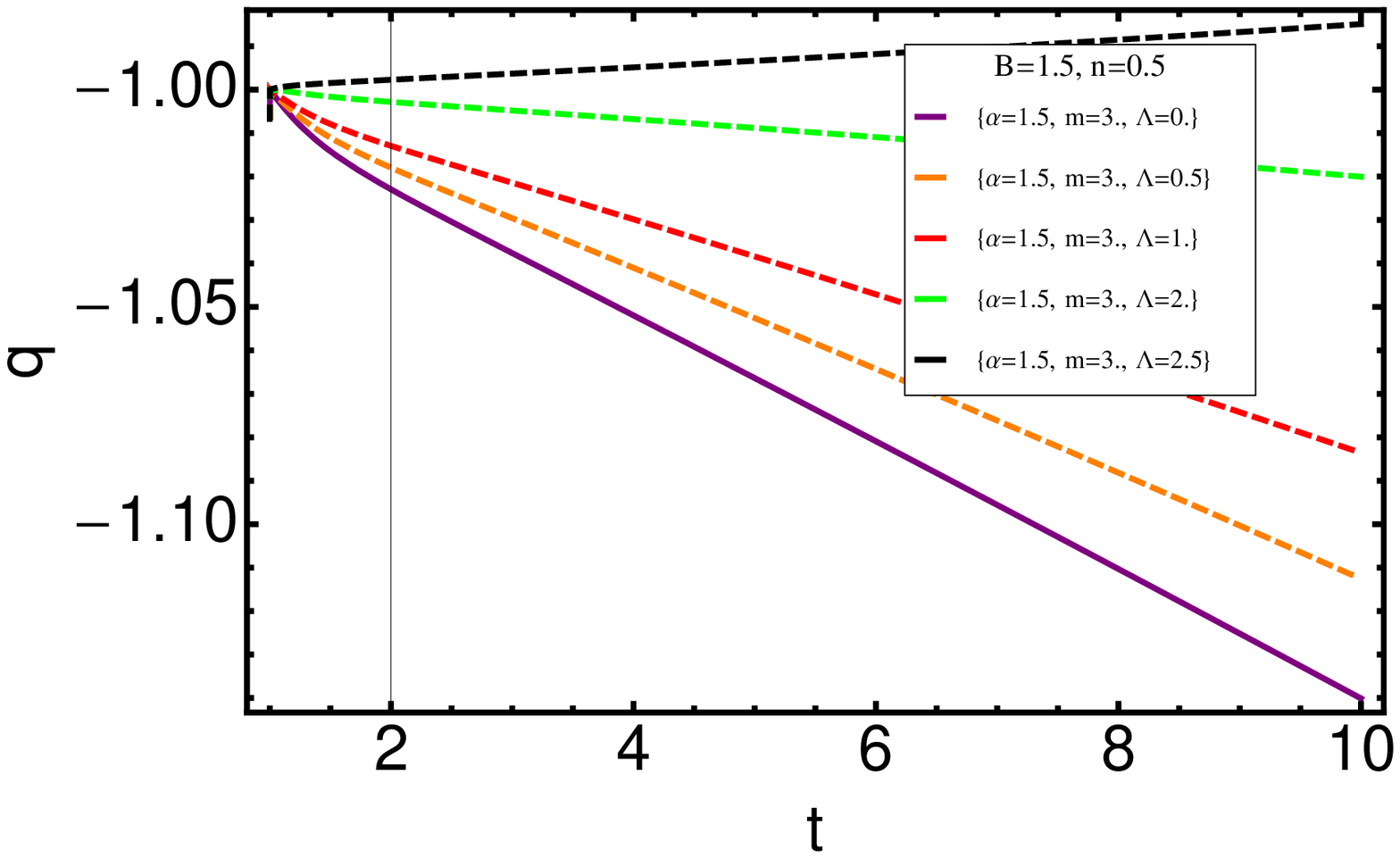}&
 \end{array}$
 \end{center}
\caption{Behavior of $H$, $\omega$ and $q$ against $t$.}
 \label{fig:3}
\end{figure}

The analysis of the second model given by Eq. (\ref{12}) is based on the following EoS parameter:
\begin{equation}\label{17}
\omega= \frac{P}{\rho} =\sum_{k=1}^{m}{A_{k}\rho^{k-1}}-\frac{B}{\rho^{n+1}}.
\end{equation}
In the plots of Fig. \ref{fig:3} we represent behavior of some cosmological parameters by numerical analysis
up to third order ($m = 3$). We obtained similar results with the first model. Therefore, we should apply a modification to obtain results which are in agreement with the current observations. For this reason, we will consider time varying $G$ and $\Lambda$ in the next section.

\section{Numerical results with varying $G$ and $\Lambda$}
To perform an analysis of the dynamics of the universe, we assume a specific model where effectively the form $\Lambda$ is given by the following relation:
\begin{equation}\label{18eq:lambda}
\Lambda=\gamma \rho,
\end{equation}
where $\gamma$ is a positive constant. Therefore, using the expression given in Eq. (\ref{7}), we have the following expression for the the dynamics of $G$:
\begin{equation}\label{19}
\dot{G} +\frac{ \gamma \dot{\rho} }{8\pi \rho}=0.
\end{equation}
Eq. (\ref{19}) can be easily integrated leading to the following expression for $G$:
\begin{equation}\label{20eq:lambda2}
G=G_{0}-\frac{\gamma}{8 \pi} \ln{\rho},
\end{equation}
where $G_{0}$ is an integration constant. In the following subsections, we give a numerical analysis of the two models.
\subsection{\large{Variable modified Chaplygin gas model}}
\begin{figure}[h!]
 \begin{center}$
 \begin{array}{cccc}
\includegraphics[width=55 mm]{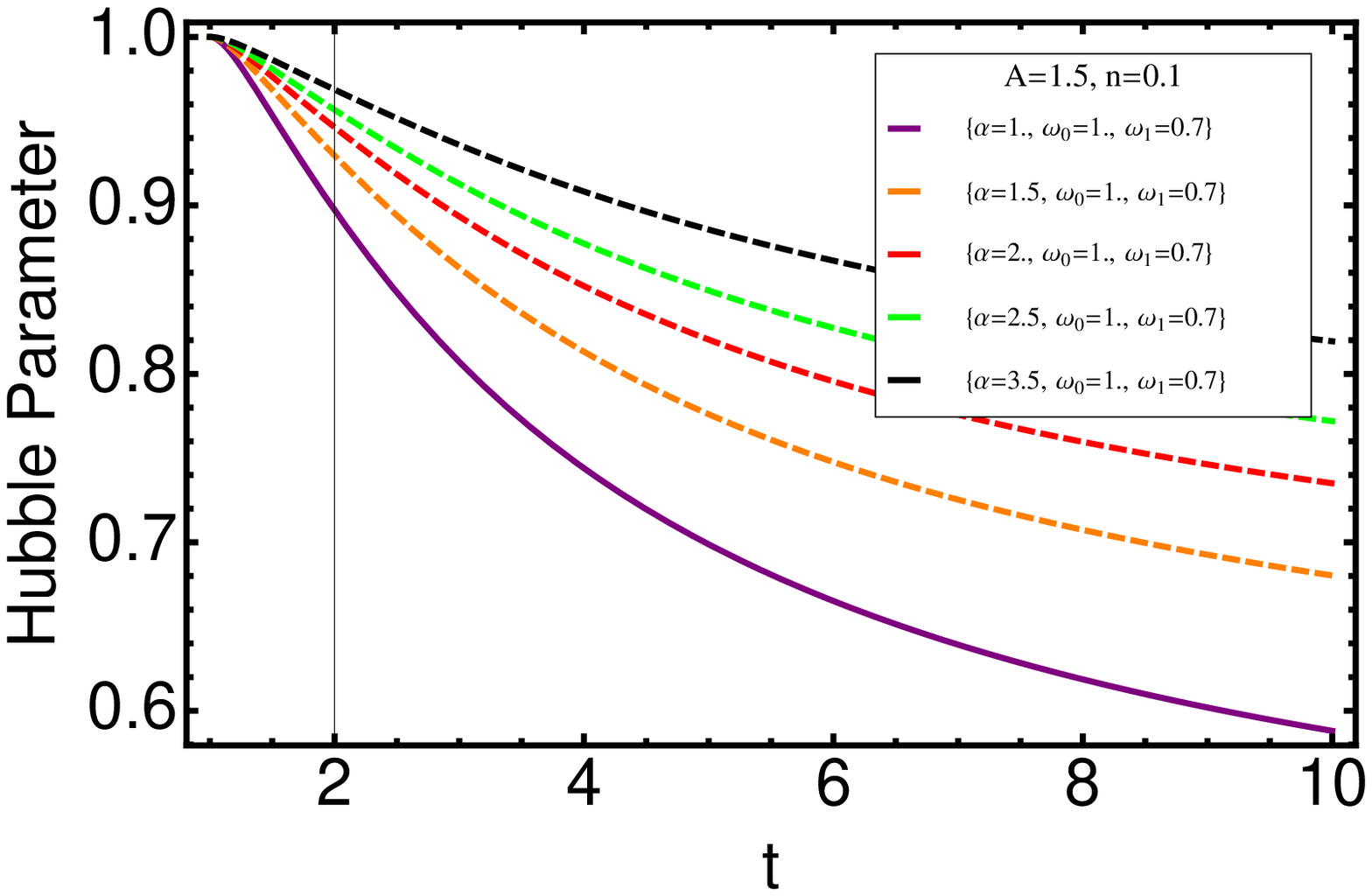} &
\includegraphics[width=55 mm]{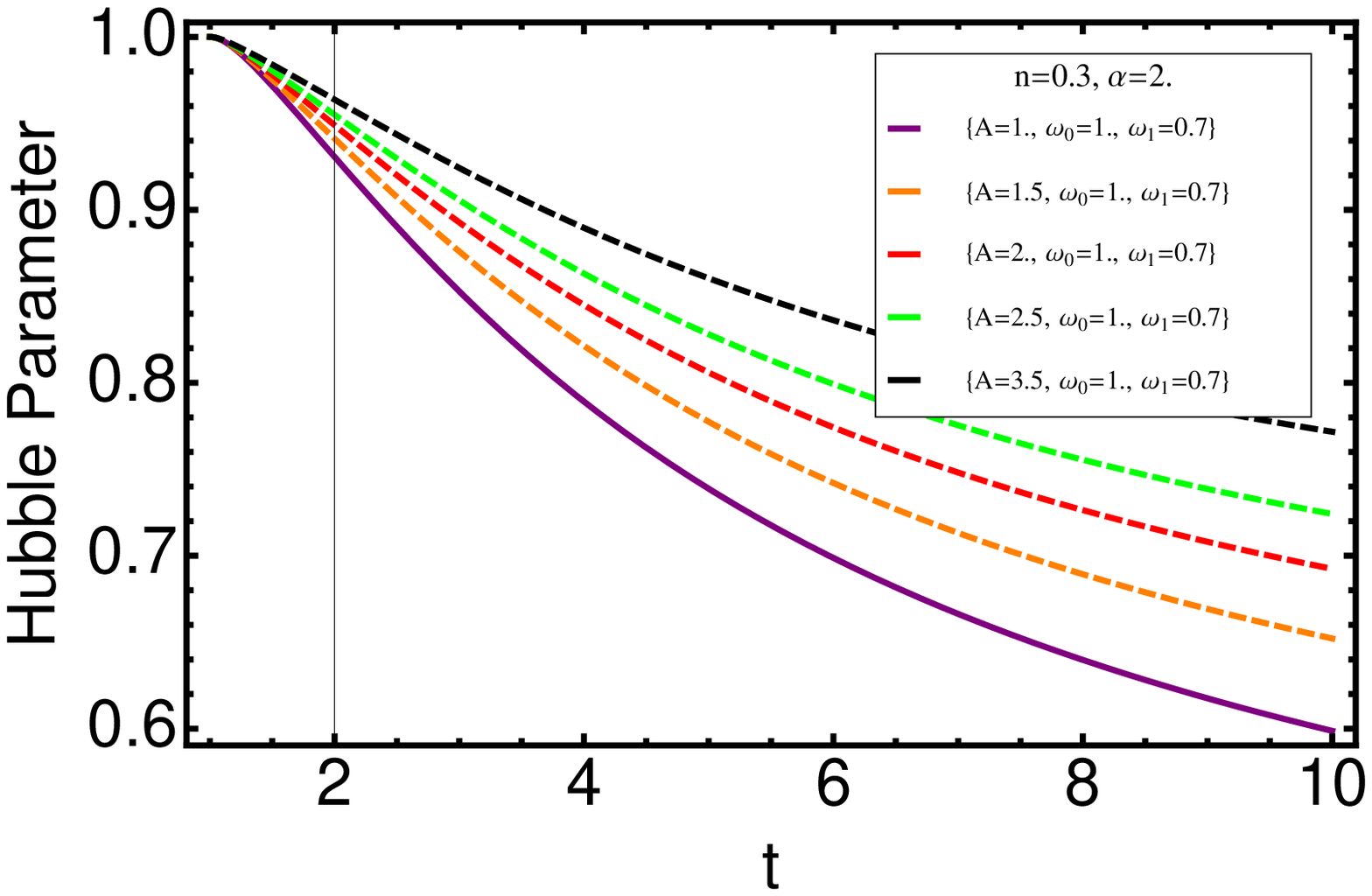}\\
\includegraphics[width=55 mm]{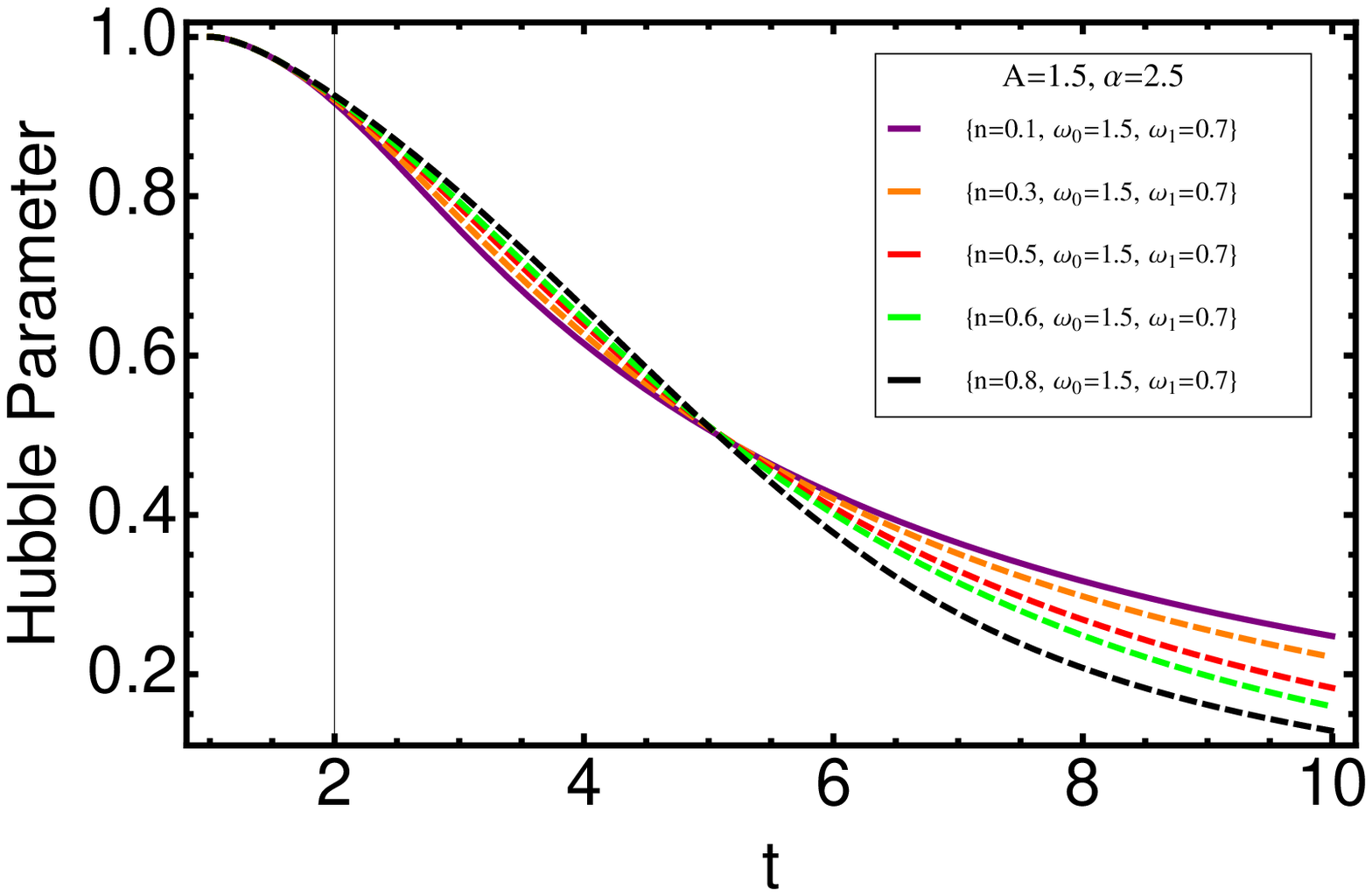}&
\includegraphics[width=55 mm]{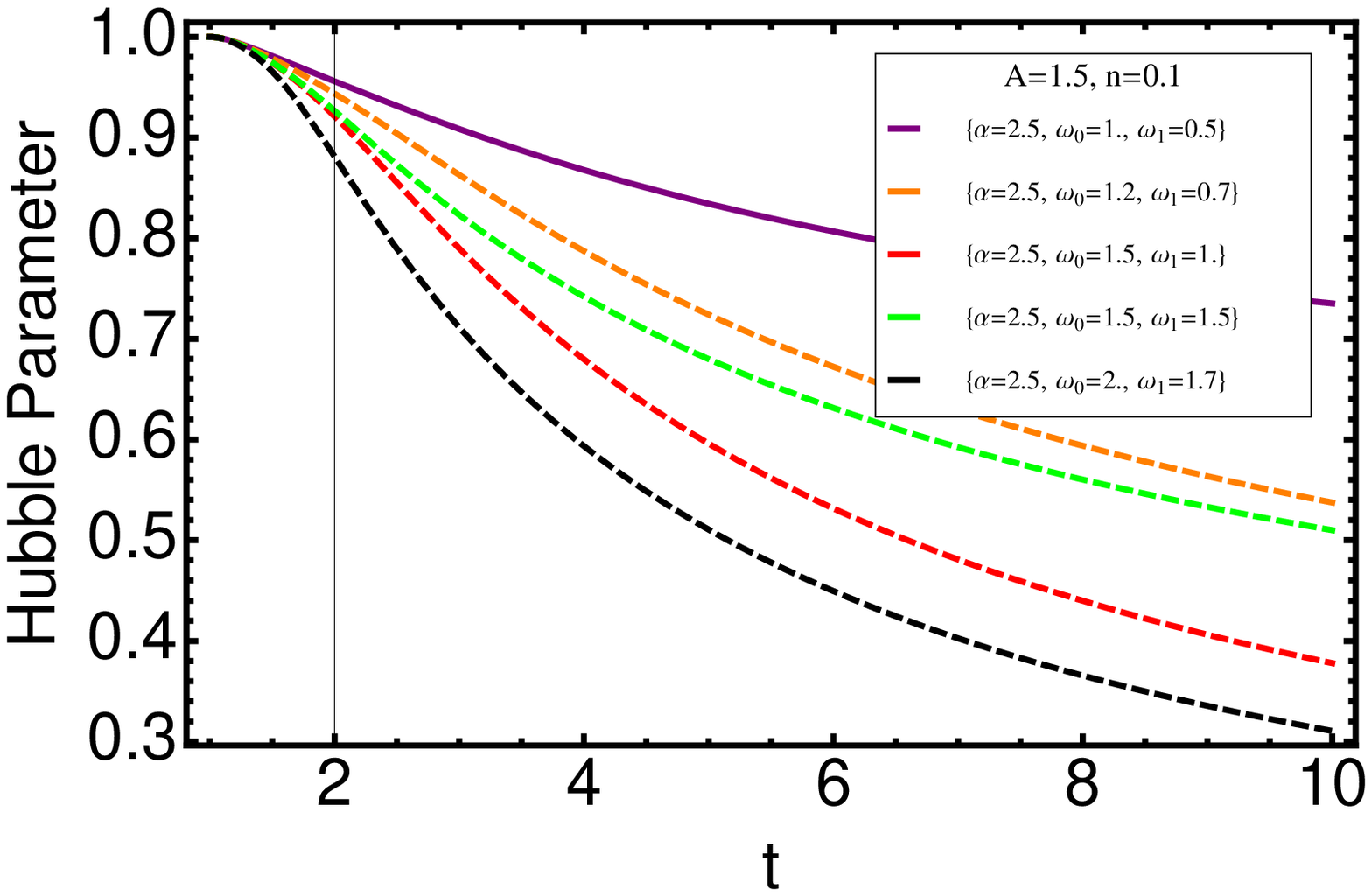}
 \end{array}$
 \end{center}
\caption{Behavior of $H$ against $t$. Model 1}
 \label{fig:4}
\end{figure}

\begin{figure}[h!]
 \begin{center}$
 \begin{array}{cccc}
\includegraphics[width=55 mm]{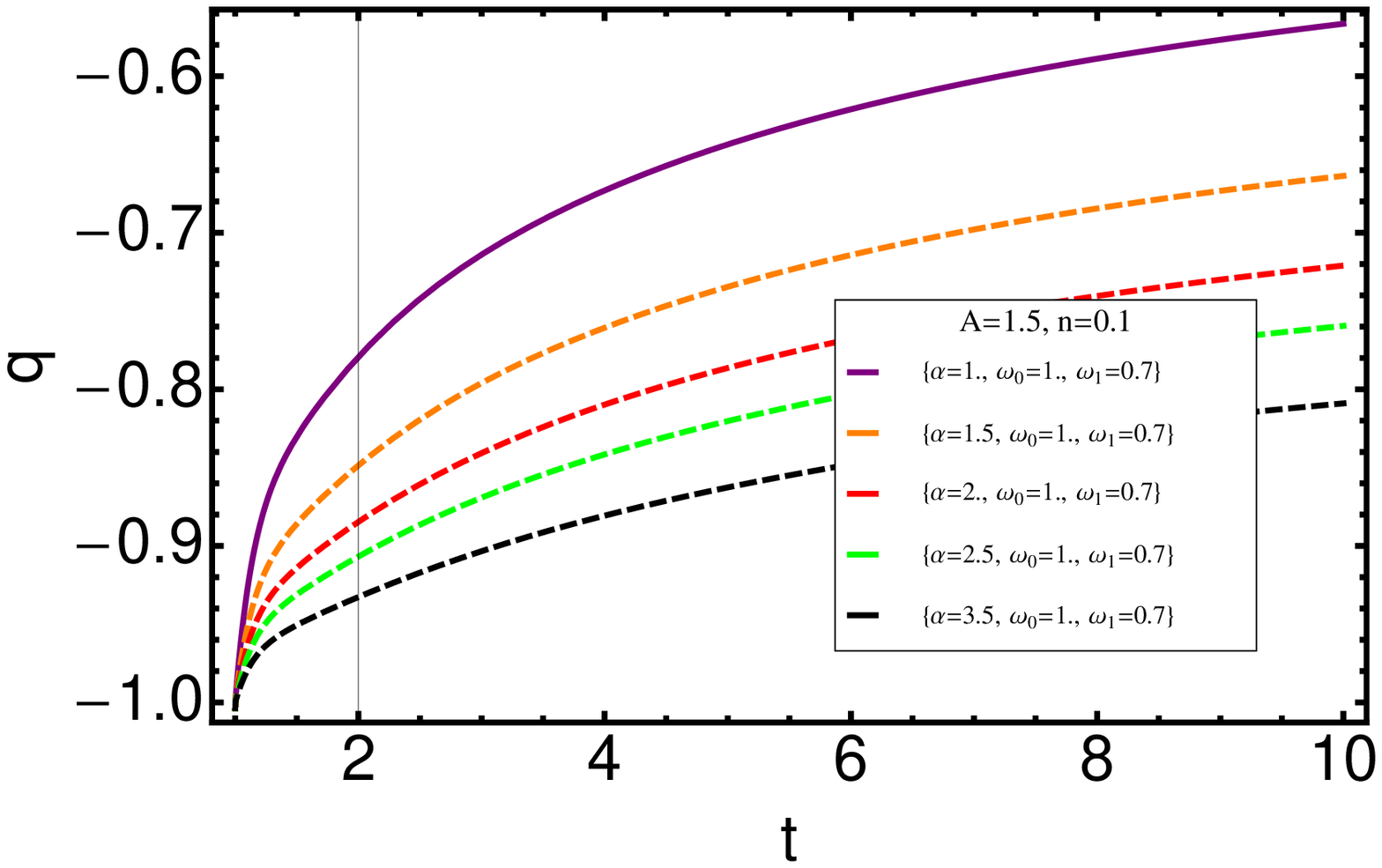} &
\includegraphics[width=55 mm]{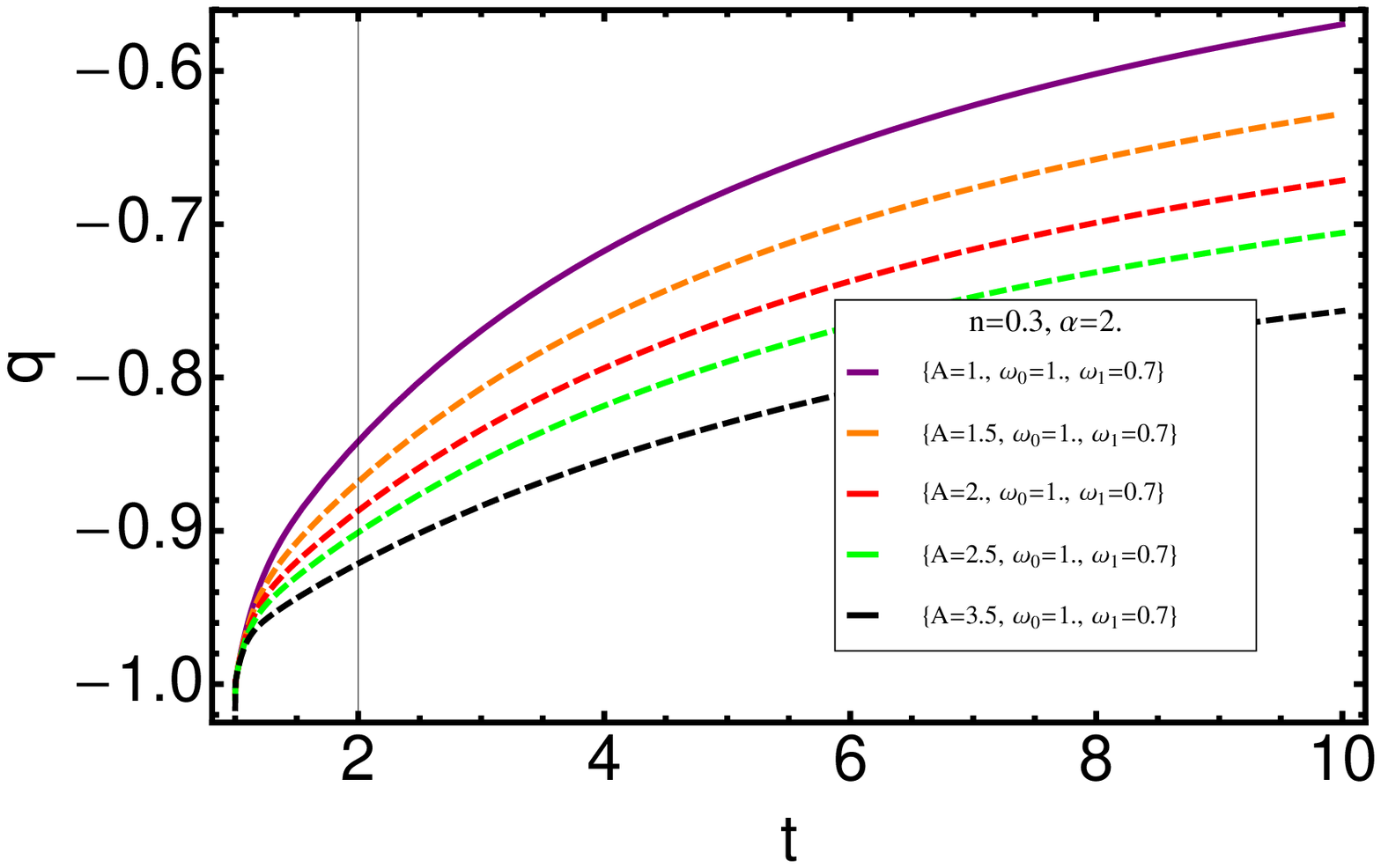}\\
\includegraphics[width=55 mm]{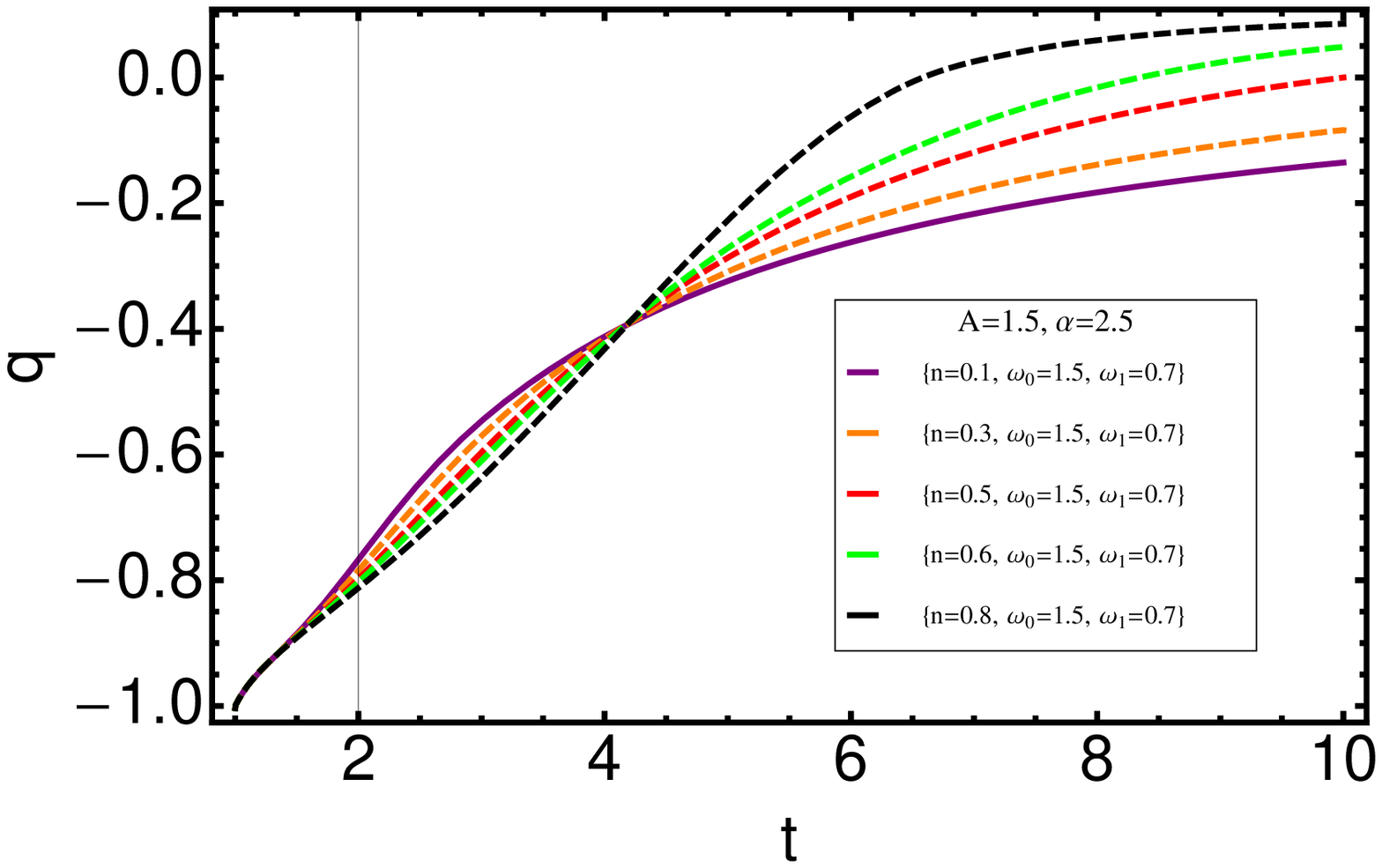}&
\includegraphics[width=55 mm]{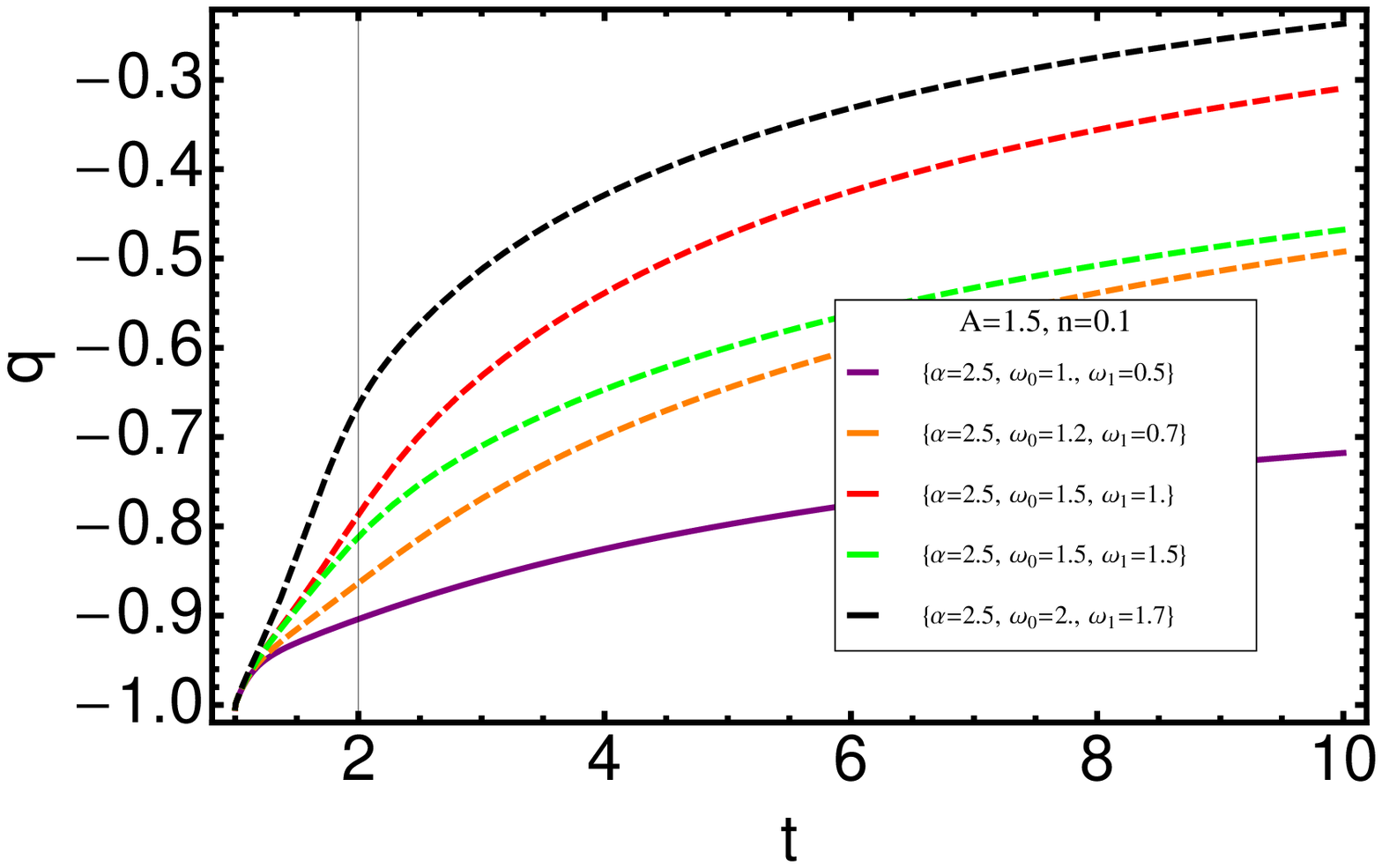}
 \end{array}$
 \end{center}
\caption{Behavior of $q$ against $t$. Model 1}
 \label{fig:5}
\end{figure}

\begin{figure}[h!]
 \begin{center}$
 \begin{array}{cccc}
\includegraphics[width=55 mm]{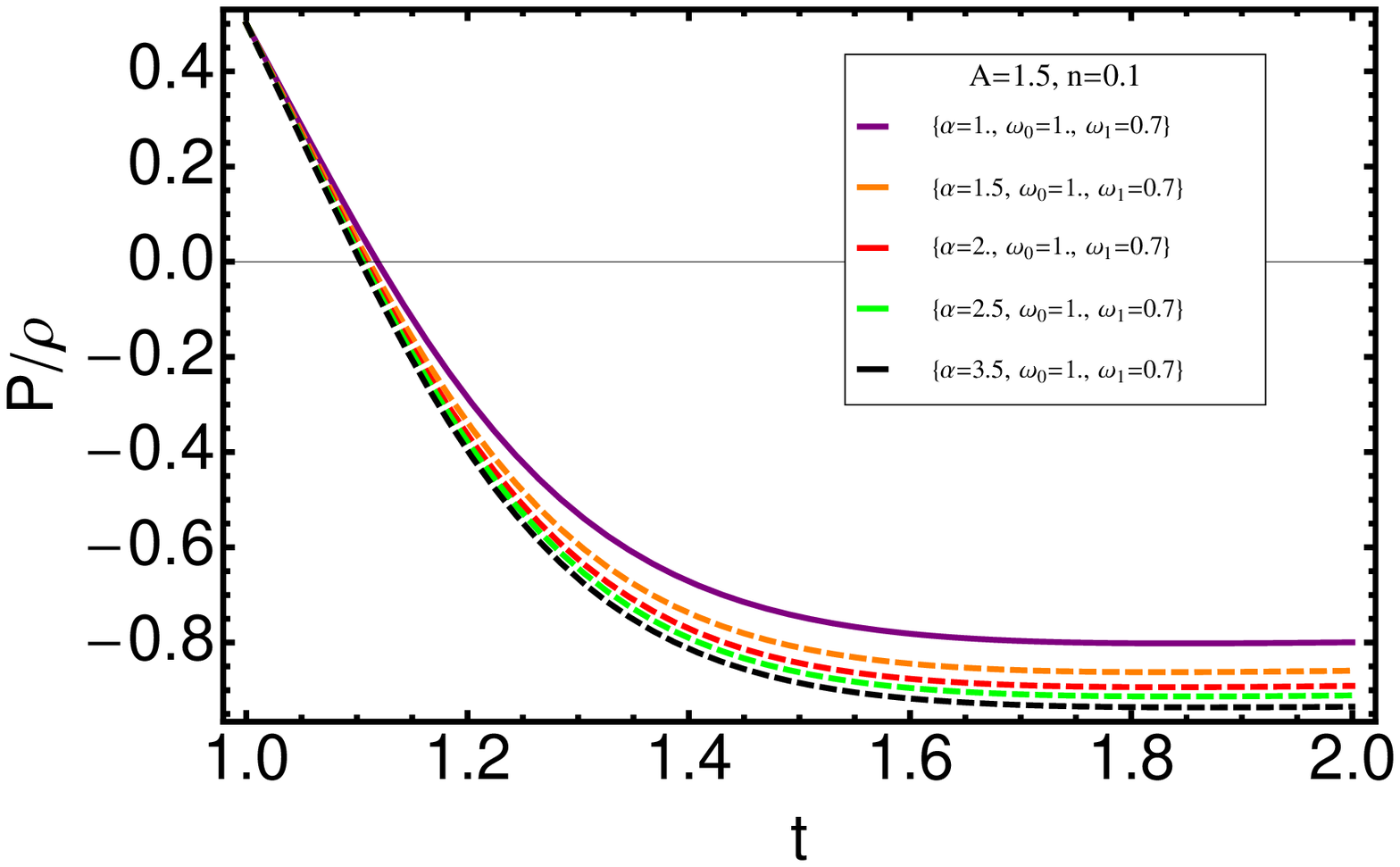} &
\includegraphics[width=55 mm]{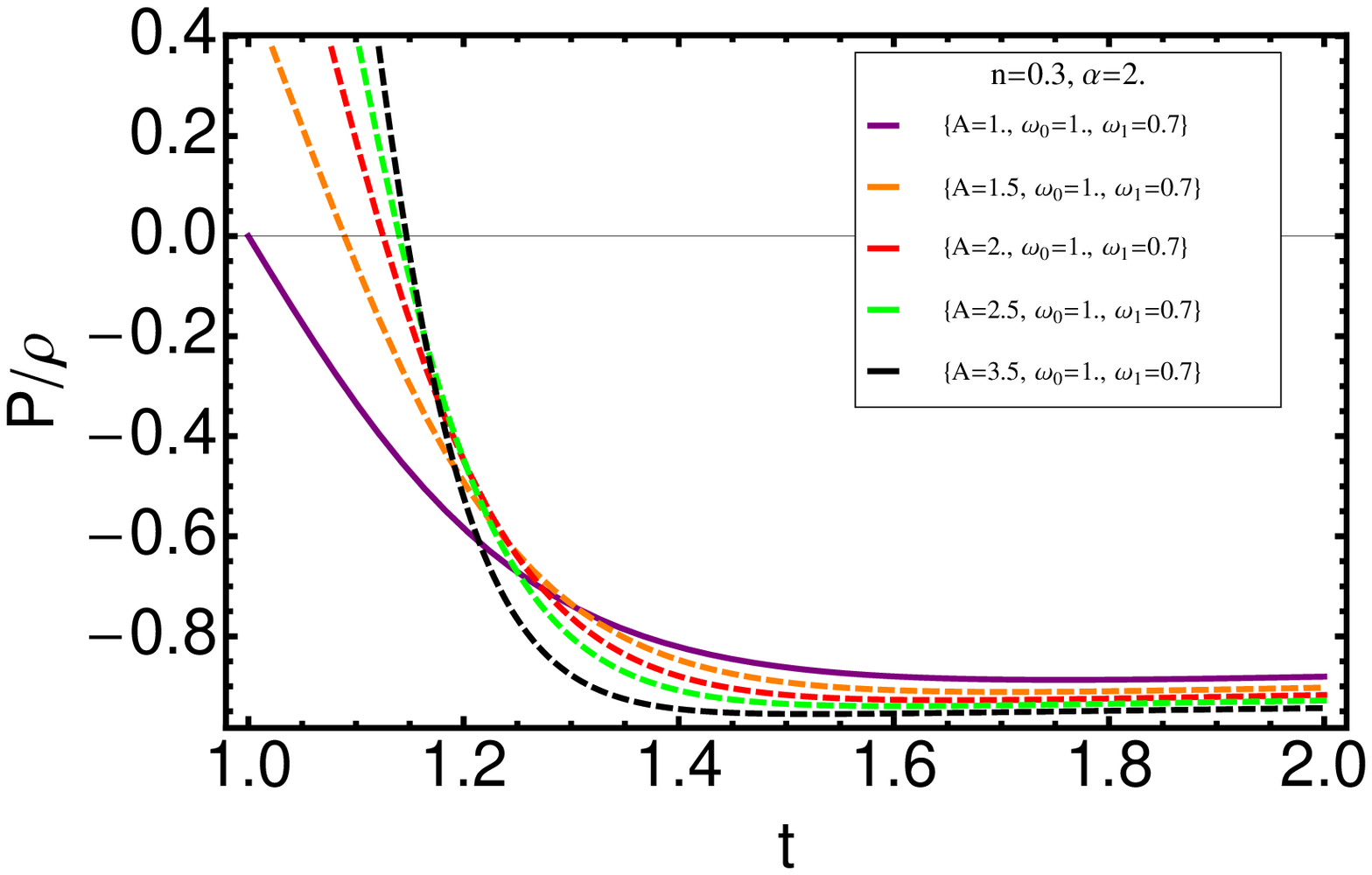}\\
\includegraphics[width=55 mm]{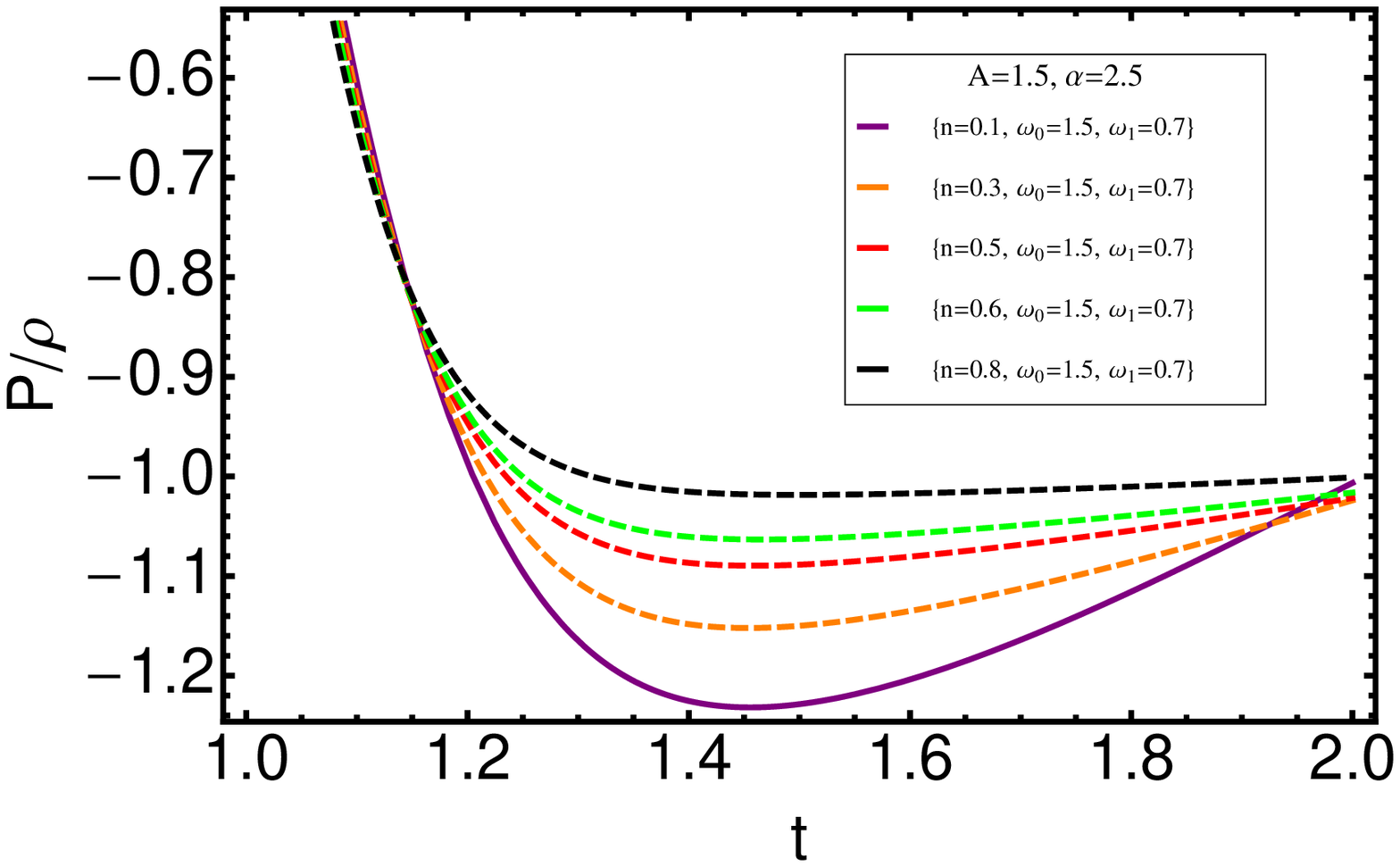}&
\includegraphics[width=55 mm]{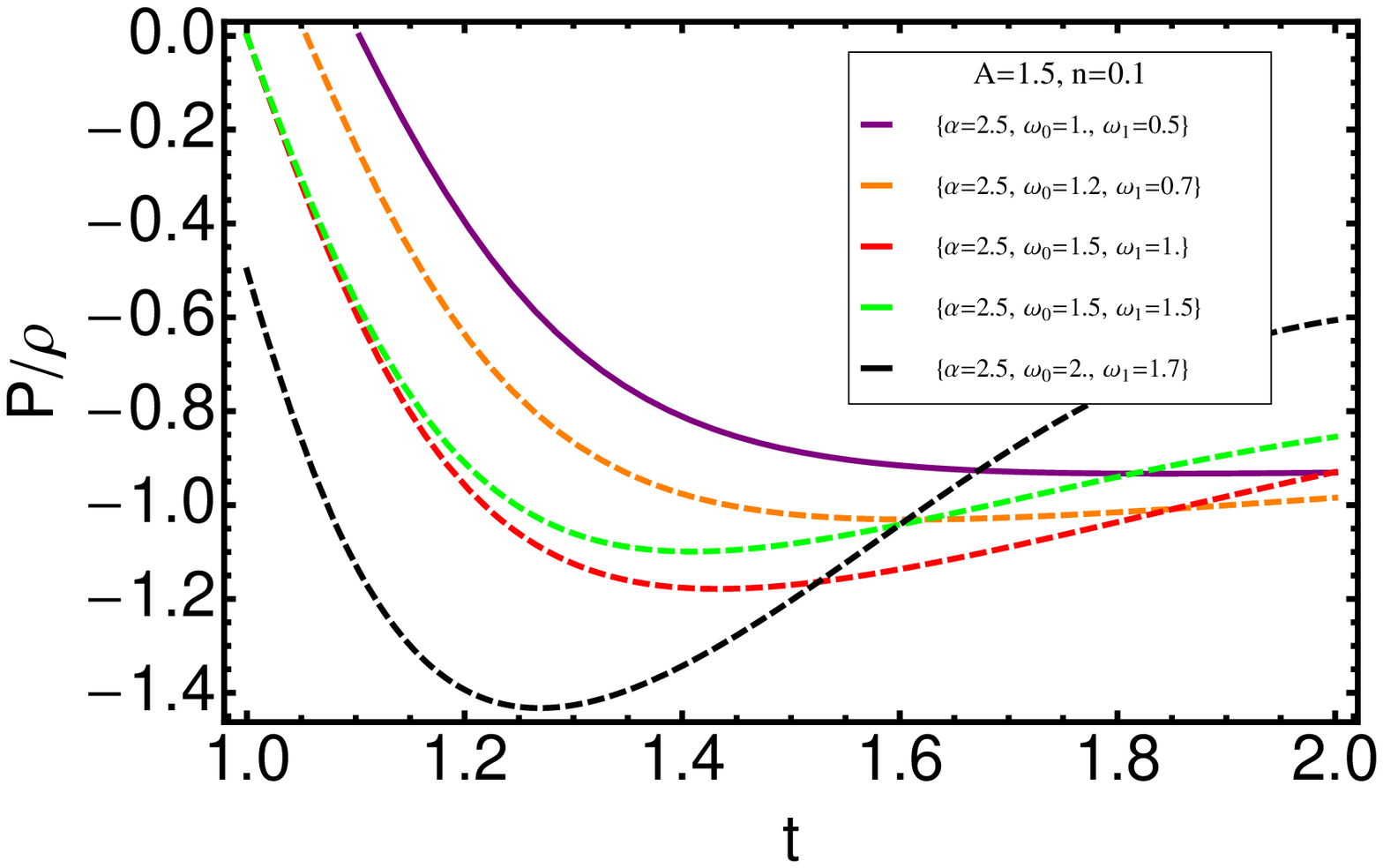}
 \end{array}$
 \end{center}
\caption{Behavior of $\omega$ against $t$. Model 1}
 \label{fig:6}
\end{figure}

Within this subsection, we analyze our first model, which is the time-varying modified Chaplygin gas with EoS given by:
\begin{equation}\label{21}
\omega=A-\frac{B(t)}{\rho^{n+1}},
\end{equation}
where we used Eq. (\ref{9}).\\
The plots of Fig. \ref{fig:4} show that the Hubble parameter in this model is a decreasing function of time
The first plot shows that increasing $\alpha$ increases the value of $H$;
therefore we find that higher order terms of gravity increases the value of Hubble parameter.
The second plot deals with the variation of parameter $A$. It has been shown that increasing $A$
increases the value of $H$. In the third plot we look at how Hubble parameter changes with $n$.
It is clear that the variation of $H$ depends on the time period. During the period when $t < 5$,
increasing $n$ increases the value of $H$; but for $t > 5$, increasing $n$ decreases the value of $H$.
Finally, in the last plot we can see the variation of $H$ with $\omega_0$ and $\omega_1$.
In the plots of Fig. \ref{fig:5} we can see the evolution of the deceleration parameter with various values of
$\alpha$, $A$, $n$, $\omega_0$ and $\omega_1$. In all cases we can see that $q$ takes value between -1 and 0,
which is in agreement with current observational data. So in this case there is no acceleration to deceleration phase transition.\\
In Fig. 6 we draw EoS parameter versus time. We can see from the first plot that
increasing $\alpha$ decreases the value of $\omega=P/\rho$.
It is illustrated that higher values of $\alpha$ makes $\omega\rightarrow-1$ faster than lower values of $\alpha$.
Similar situation happens by varying $A$ (at least for the late time behavior).
The third plot of Fig. \ref{fig:6} shows the variation of $\omega$ with $n$.
The first and the second plots were drawn for $n = 0.1$ and $n = 0.3$ respectively causing $\omega$ to
be a decreasing function of time and eventually asymptote to constant negative values.
But, for different values of $n$ and $\omega$ the situation is different, as illustrated in the third and
forth plot of Fig. \ref{fig:6}.

\begin{figure}[h!]
 \begin{center}$
 \begin{array}{cccc}
\includegraphics[width=55 mm]{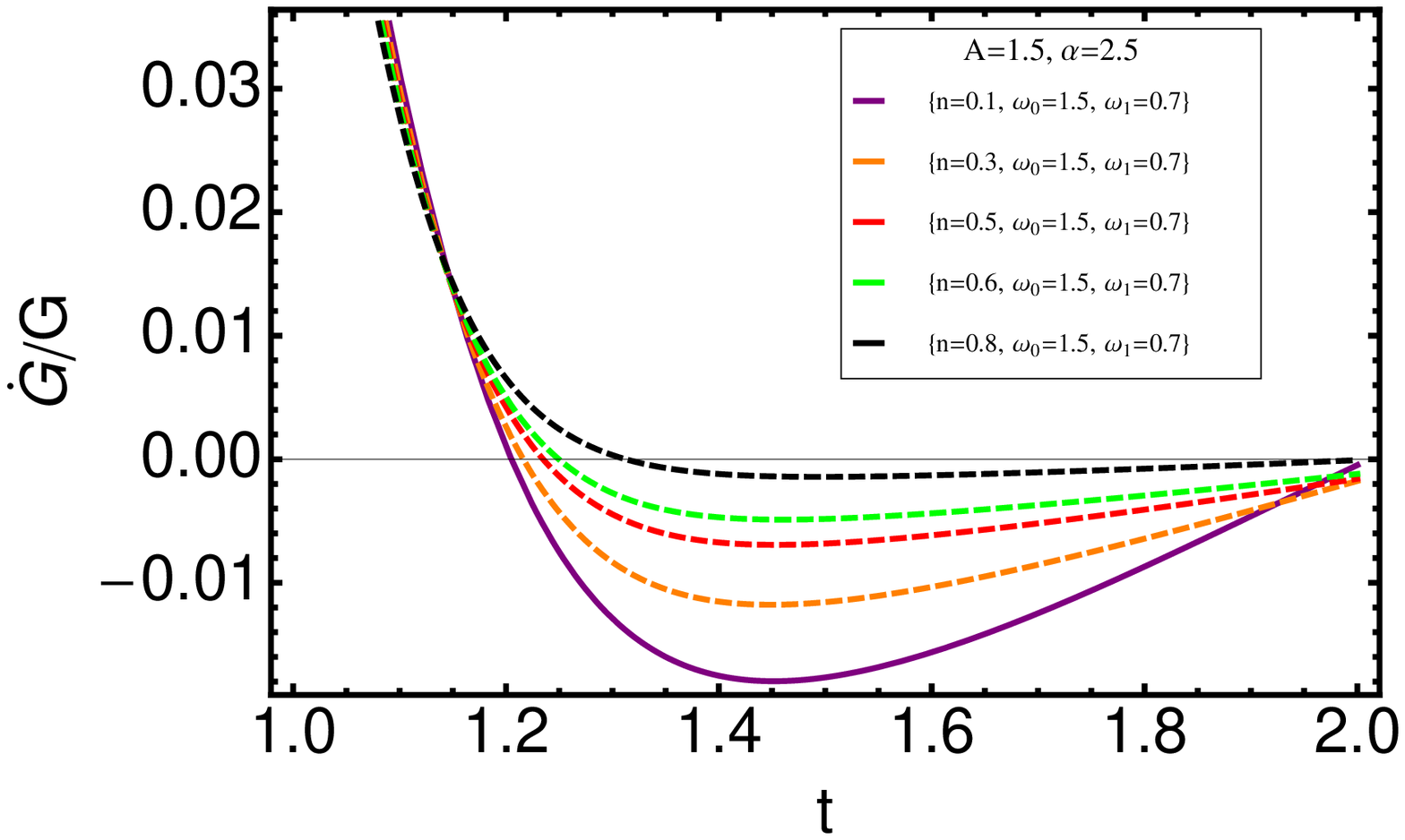}&
\includegraphics[width=55 mm]{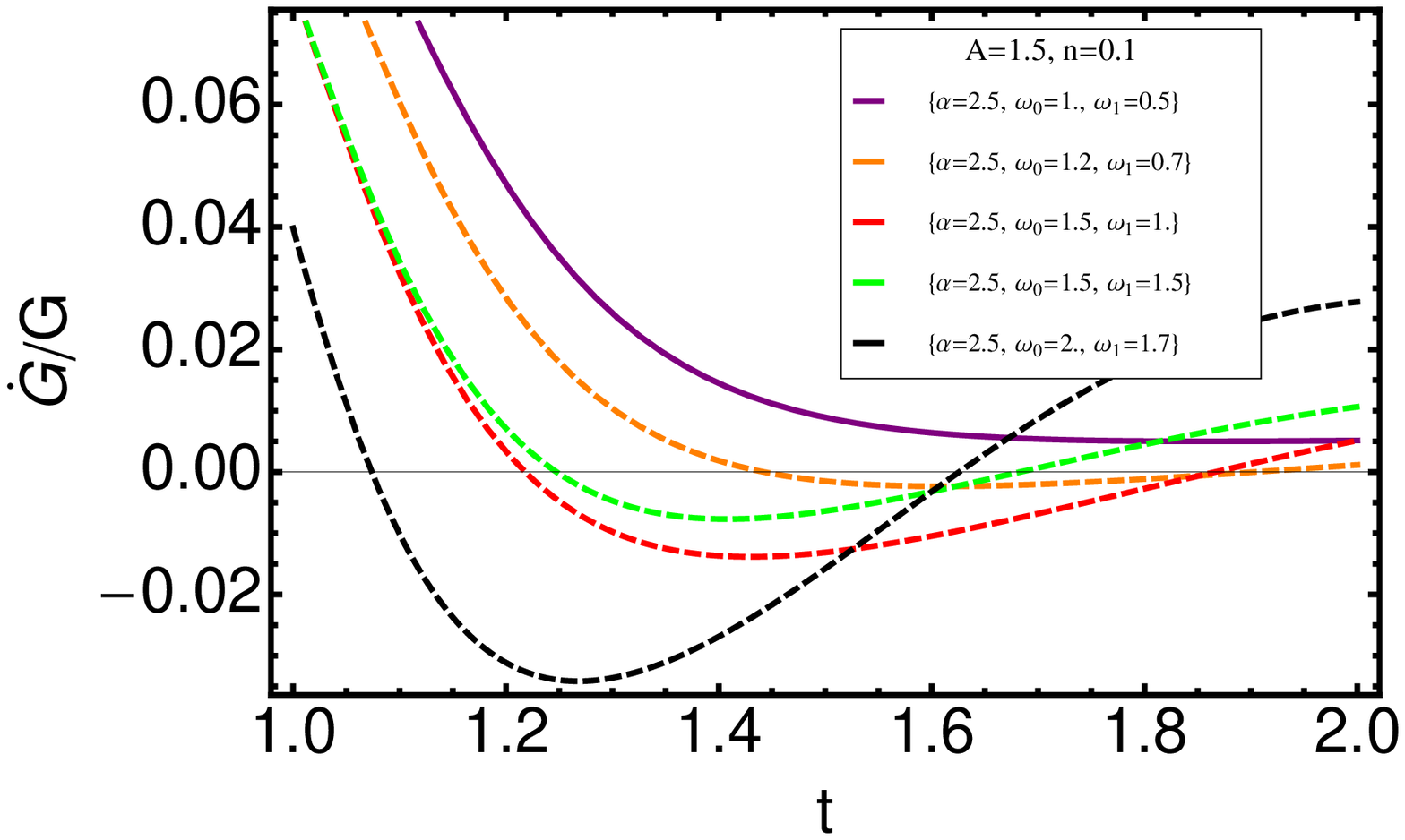}
 \end{array}$
 \end{center}
\caption{Behavior of $\dot{G}/G$ against $t$ in unit of $10^{-11}yr^{-1}$. Model 1}
 \label{fig:7}
\end{figure}

Finally, in Fig. \ref{fig:7} we draw the variation of $\dot{G}/G$ versus time. According to the most observational data, we have \cite{P42}:
\begin{equation}\label{22}
|\frac{\dot{G}}{G}|\leq1.3\times10^{-12}yr^{-1}.
\end{equation}

The plots of Fig. \ref{fig:7} show a good agreement with the observational constraint given in Eq. (\ref{22}).\\
Therefore, we can conclude that the first model agree with observational data with the exception of the results of the deceleration parameter.

\subsection{Extended Chaplygin gas model}
\begin{figure}[h!]
 \begin{center}$
 \begin{array}{cccc}
\includegraphics[width=55 mm]{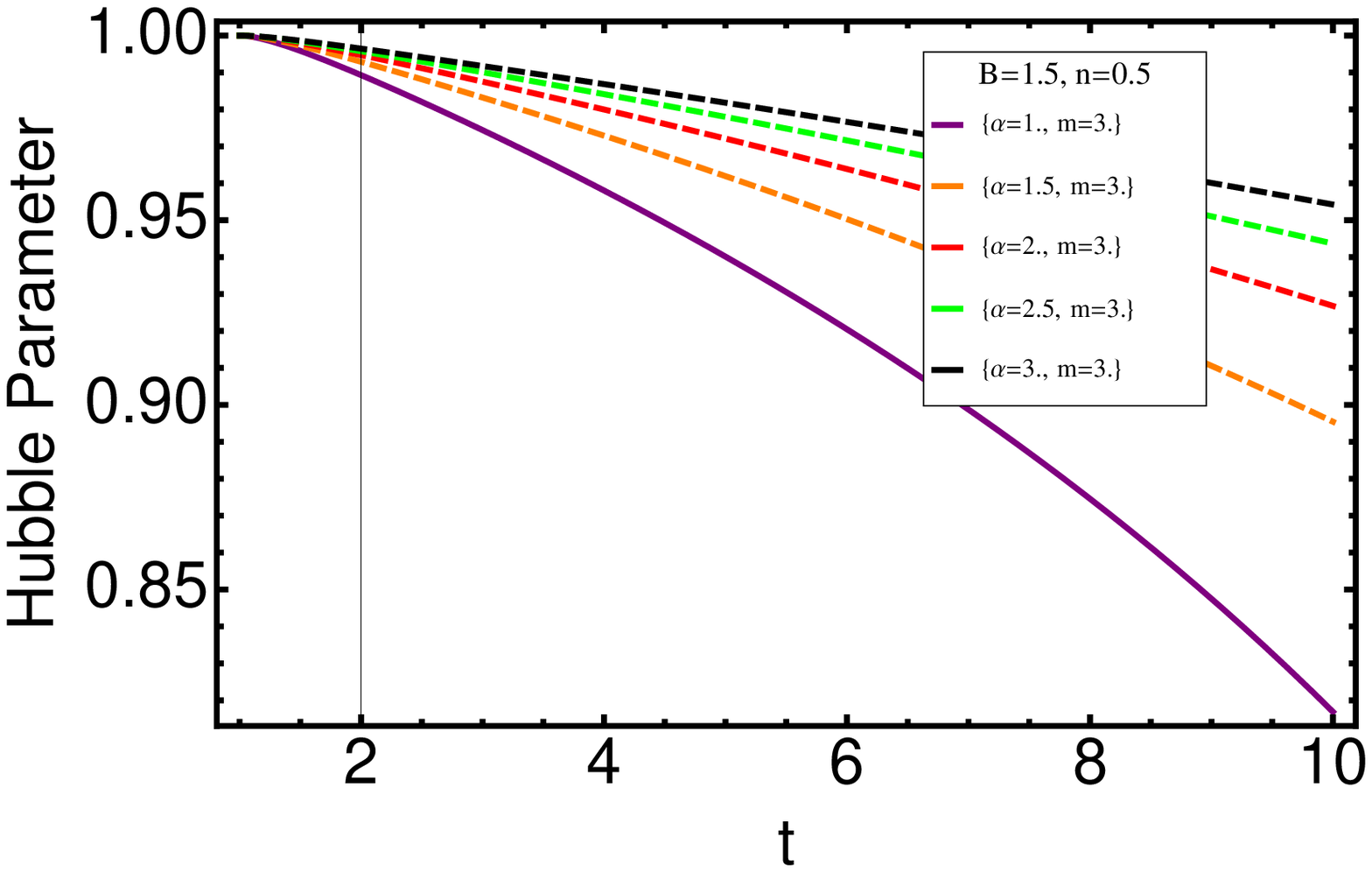} &
\includegraphics[width=55 mm]{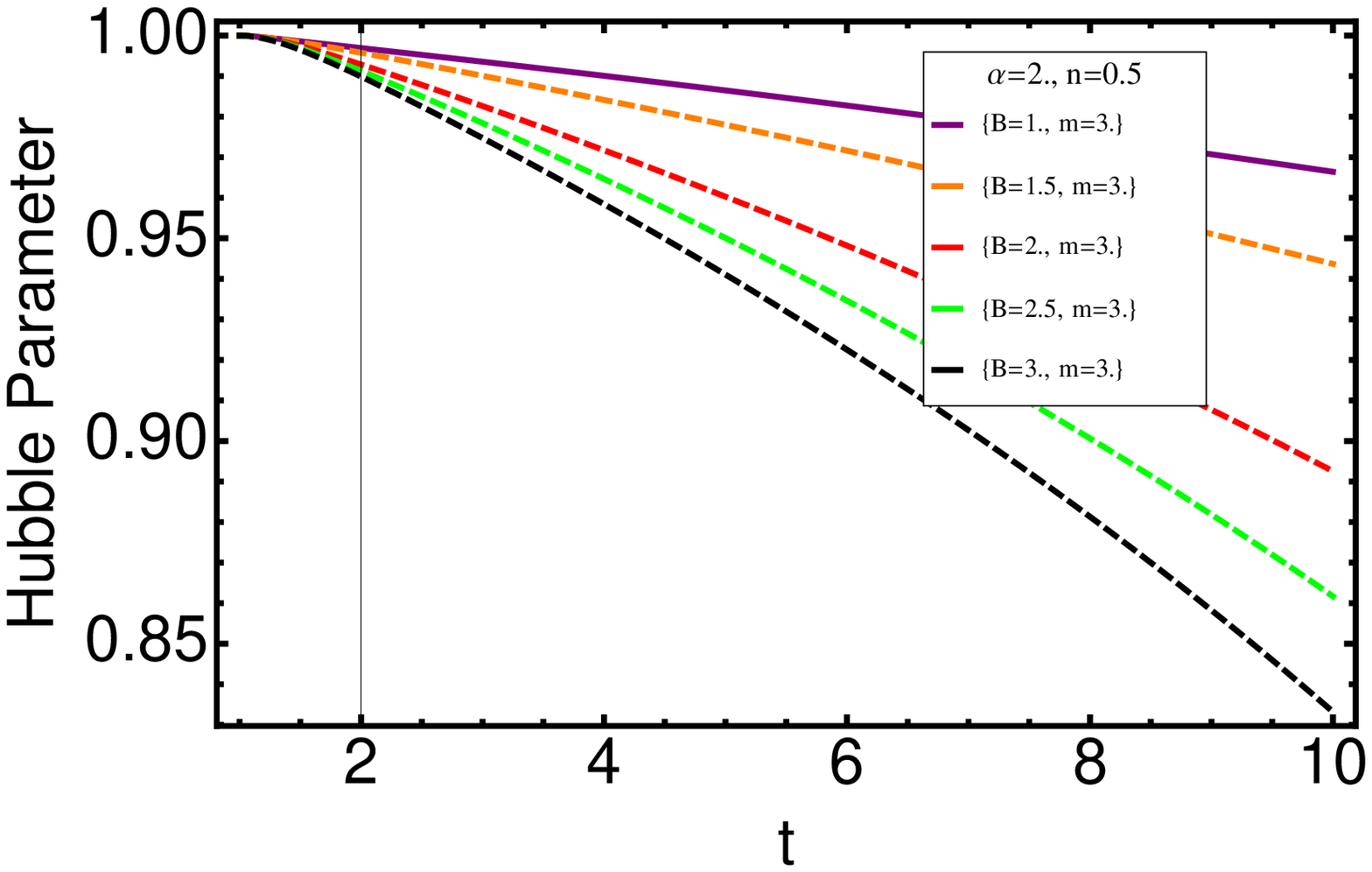}\\
\includegraphics[width=55 mm]{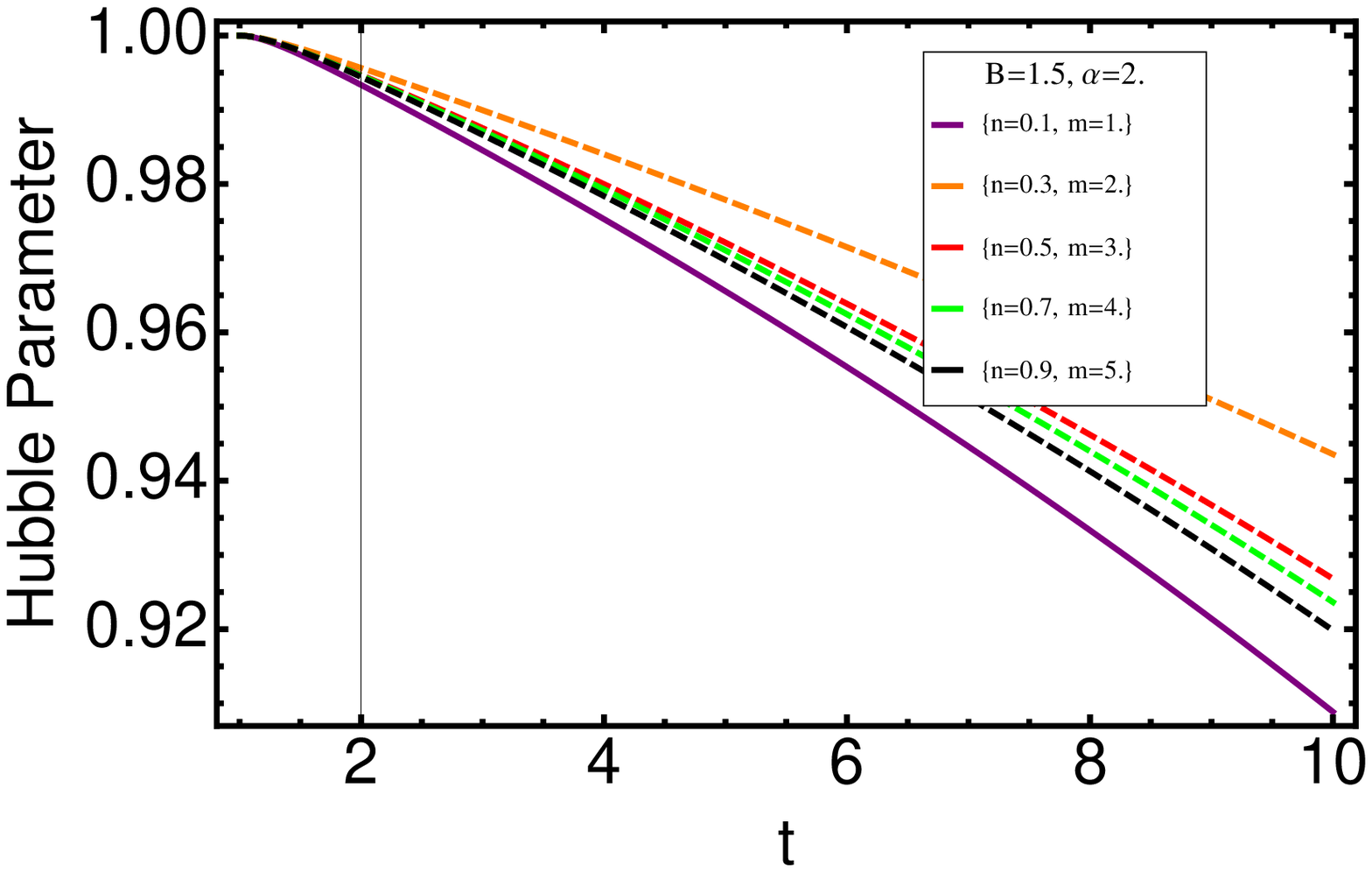}&
\includegraphics[width=55 mm]{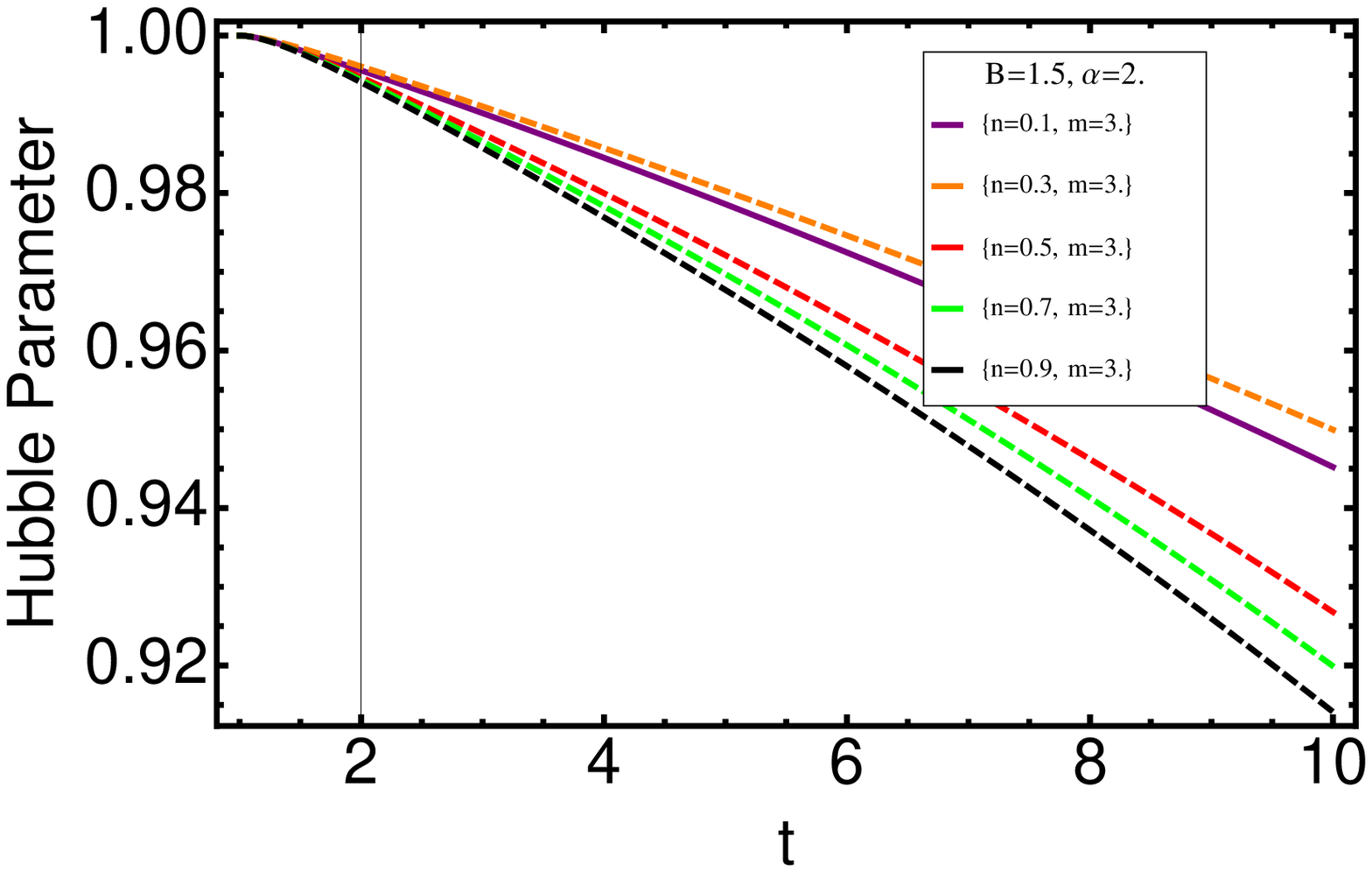}
 \end{array}$
 \end{center}
\caption{Behavior of $H$ against $t$. Model 2}
 \label{fig:8}
\end{figure}

In the second model we use the EoS parameter given in Eq. (\ref{17}). The plots of Fig. \ref{fig:8} show the behavior of Hubble parameter as a function of time.\\

\begin{figure}[h!]
 \begin{center}$
 \begin{array}{cccc}
\includegraphics[width=60 mm]{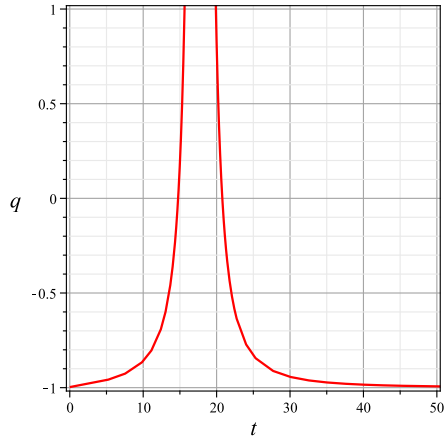} &
 \end{array}$
 \end{center}
\caption{Behavior of $q$ against $t$. Model 2}
 \label{fig:9}
\end{figure}

We can see that the Hubble parameter is a decreasing function of time. These plots suggest the following general behavior for $H$:
\begin{equation}\label{23}
H=H_{0}-Ct(1+t),
\end{equation}
where $C$ is a positive constant. Using this fit for the Hubble parameter we can obtain the following expression
for the deceleration parameter $q$:
\begin{equation}\label{16!!!}
q= -1 + \frac{C\left(1+2t\right)}{\left[ H_{0}-Ct(1+t)  \right]^2}.
\end{equation}
We plotted the behavior of the deceleration parameter obtained in Eq. (\ref{16!!!}) in Fig. \ref{fig:9}.
We can see that $q$ is negative during the early universe as well as the late universe therefore the universe
is accelerating during these two eras. And between these two epochs $q$ is positive,
hence the universe is decelerating. This behavior is in agreement with the $\Lambda$CDM model.

\begin{figure}[h!]
 \begin{center}$
 \begin{array}{cccc}
\includegraphics[width=55 mm]{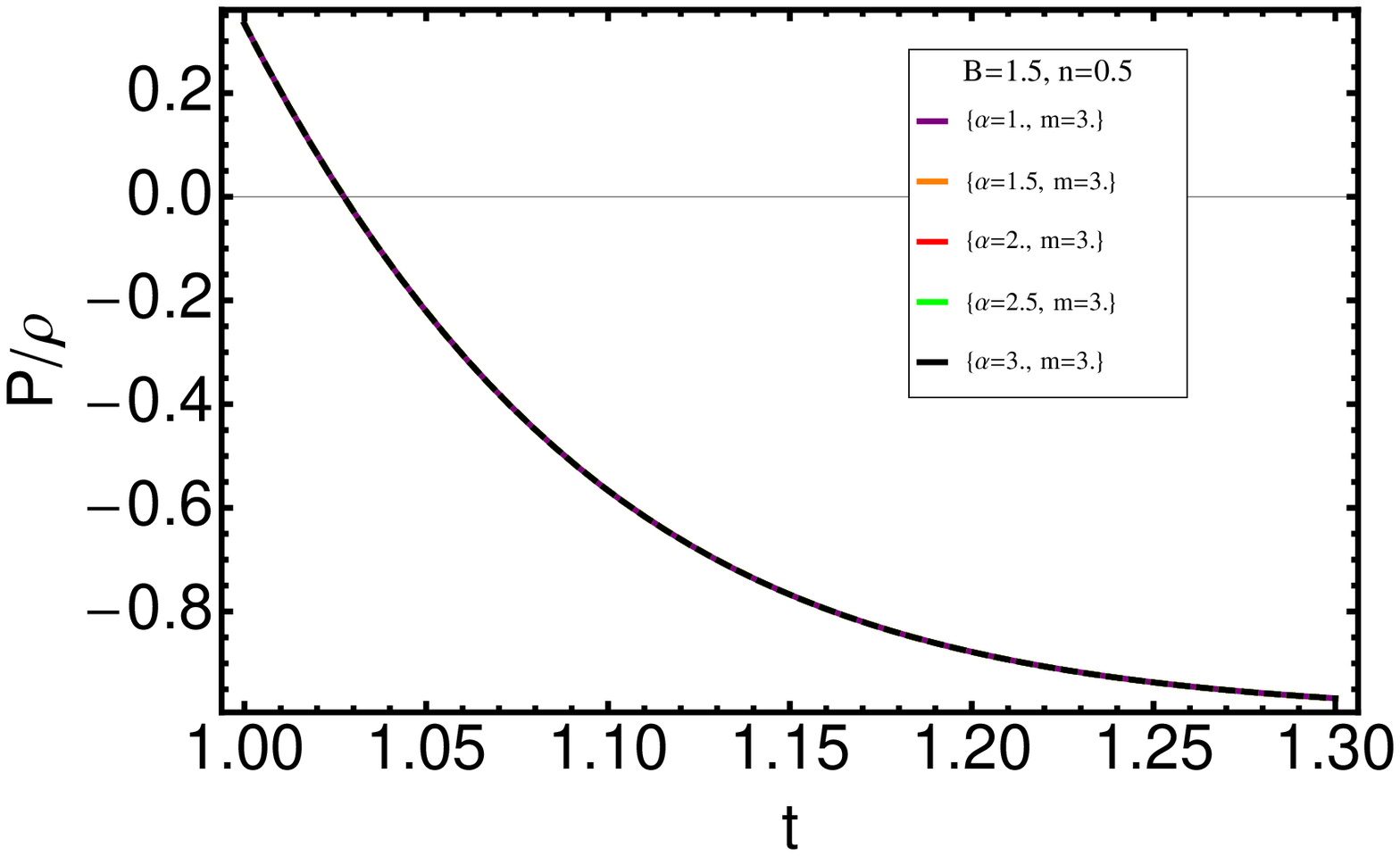} &
\includegraphics[width=55 mm]{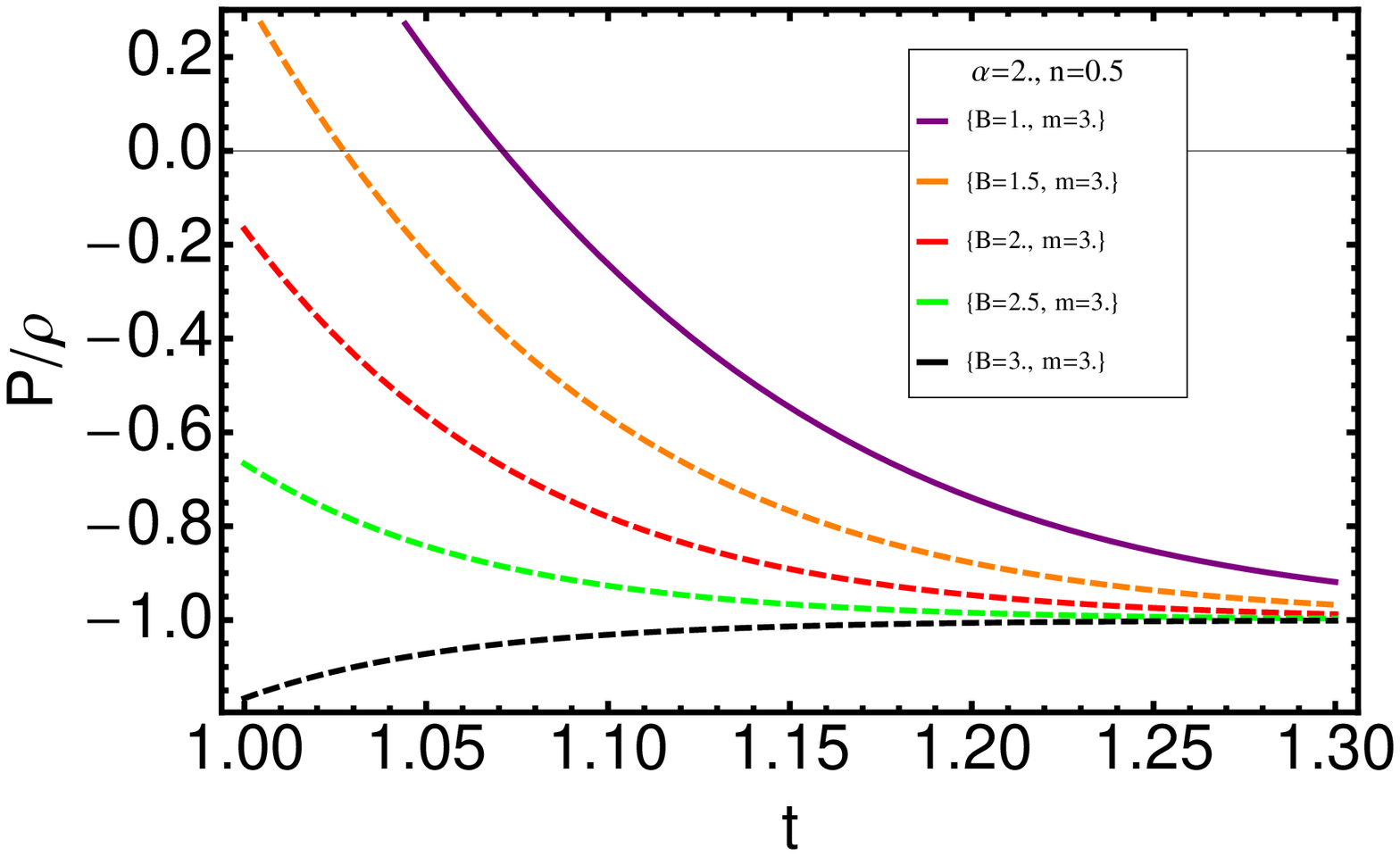}\\
\includegraphics[width=55 mm]{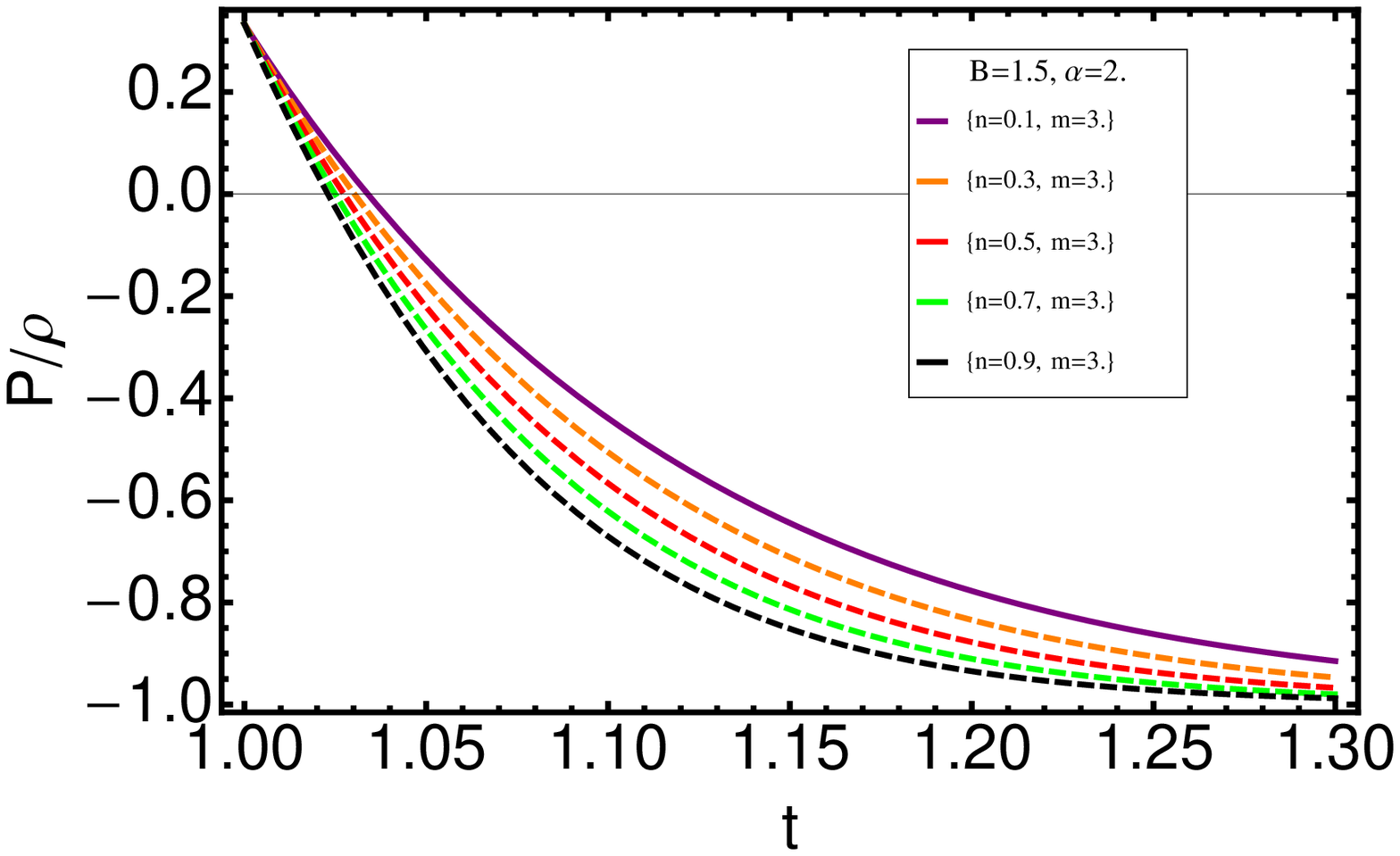}&
\includegraphics[width=55 mm]{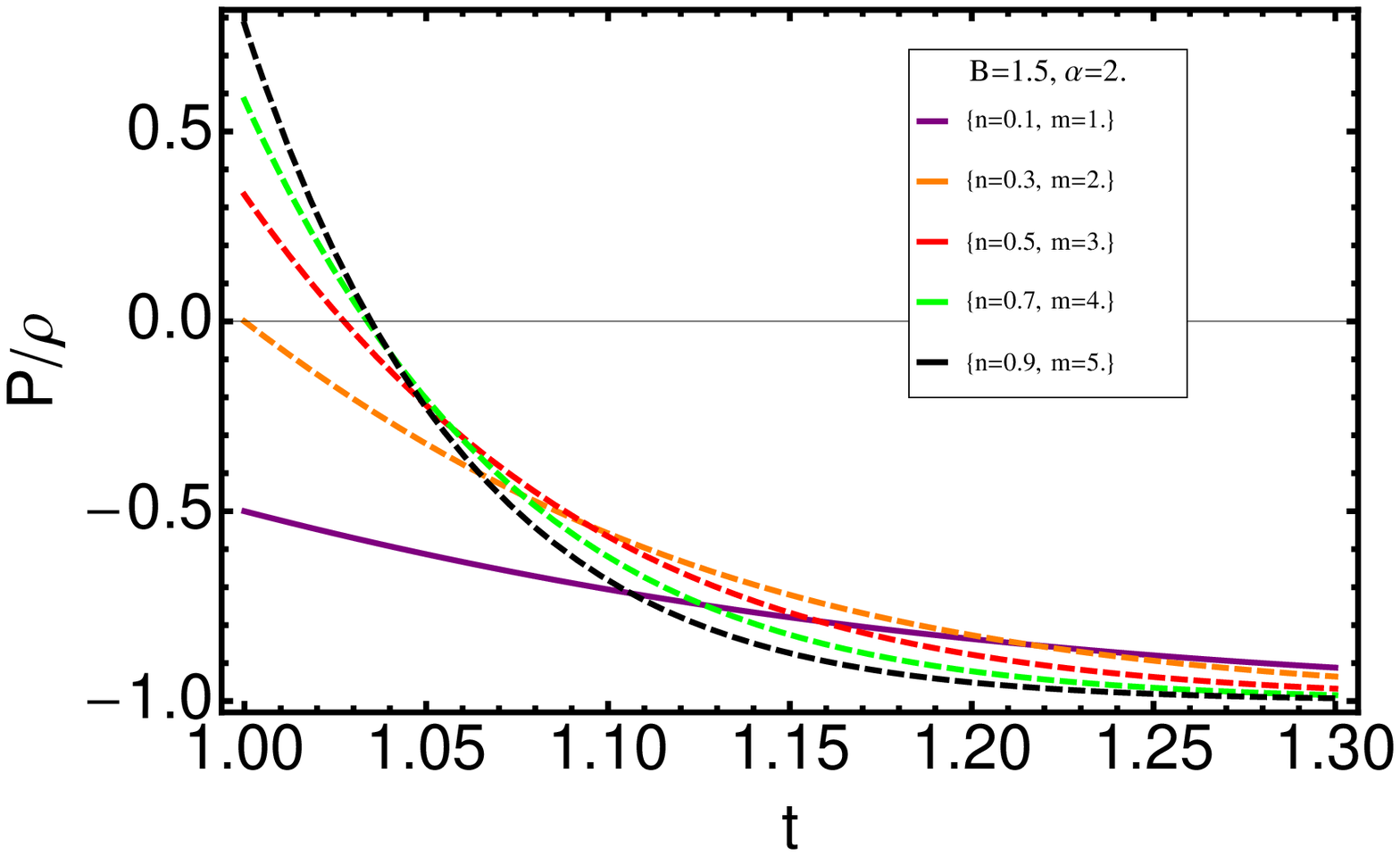}
 \end{array}$
 \end{center}
\caption{Behavior of $\omega$ against $t$. Model 2}
 \label{fig:10}
\end{figure}

\begin{figure}[h!]
 \begin{center}$
 \begin{array}{cccc}
\includegraphics[width=60 mm]{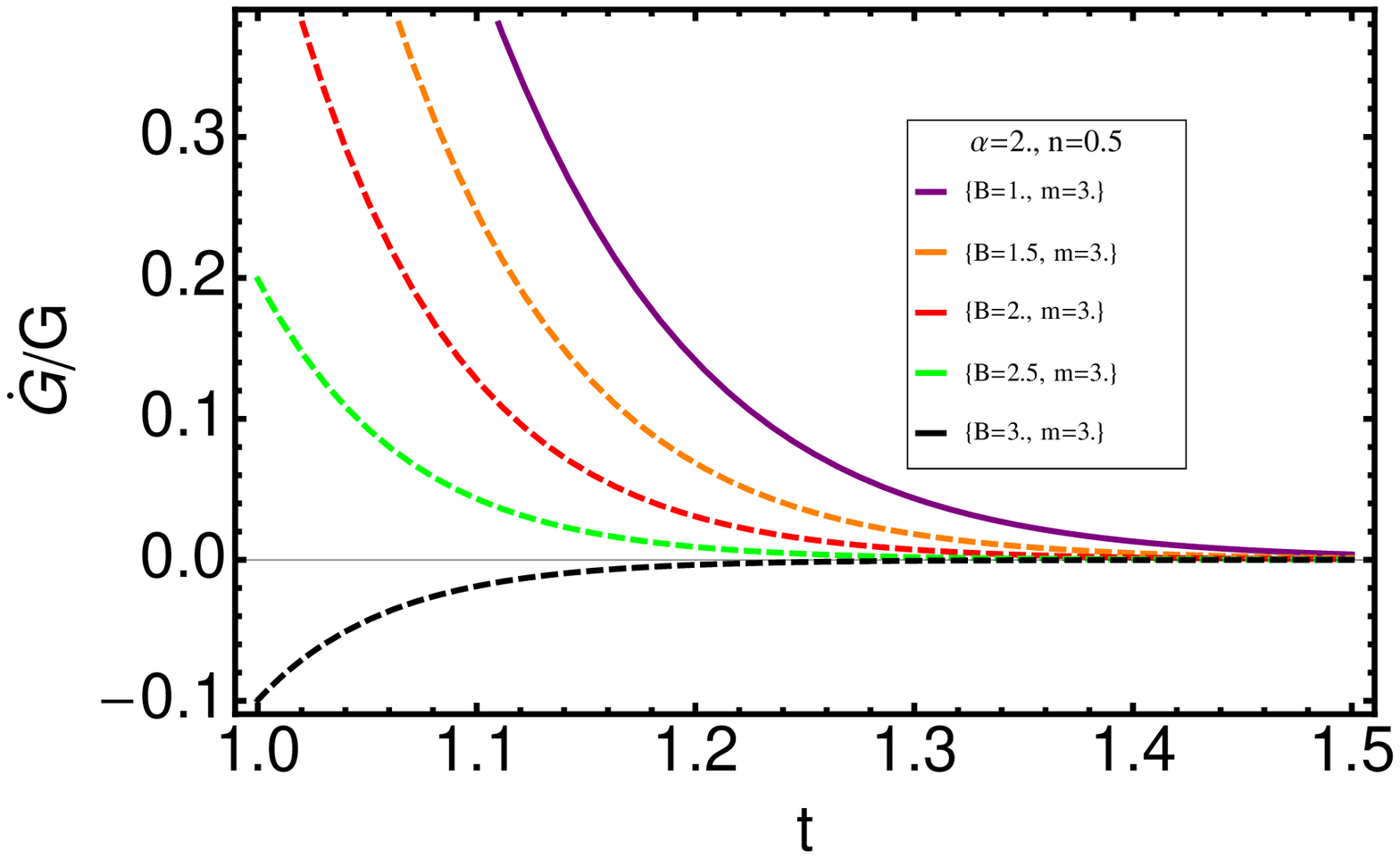}&
\includegraphics[width=60 mm]{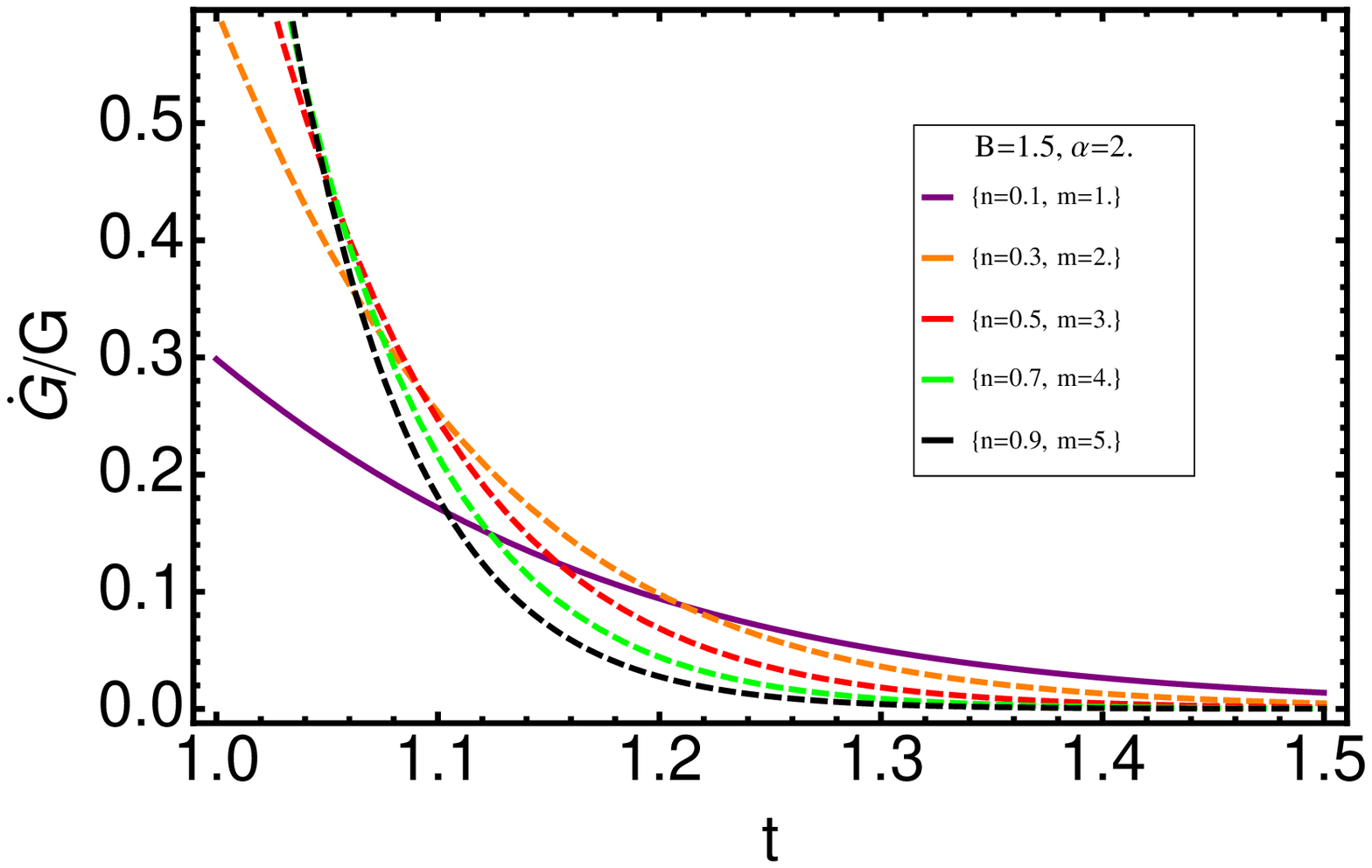}
 \end{array}$
 \end{center}
\caption{Behavior of $\dot{G}/G$ against $t$. Model 1}
 \label{fig:11}
\end{figure}

In Fig. \ref{fig:10}, we plot the EoS parameter as function of the time. The first plot shows that varying of $\alpha$ does not have an effect on $\omega$ in the presence of higher order terms. The second plot deals with the variation of the constant $B$. For the cases of $B < 3$ the EoS parameter is a decreasing function of time which tends to $-1$, while for the case of $B = 3$ the EoS parameter is an increasing function of time which also goes to the value of $-1$. The third plot shows that increasing $n$ decreases the value of $\omega$. These plots are obtained for $m = 3$. The last plot contains 5 curves corresponding to $m = 1, ..., 5$ with increasing $n$. The case of $m = 1$ which corresponds to modified Chaplygin gas is illustrated by violet line.
In the plots of Fig. \ref{fig:11} we can see the evolution of $\dot{G}/G$ in agreement with observational data.

\section{Statefinder diagnostics}
In the framework of general relativity, dark energy can explain the present cosmic acceleration. Except cosmological constant there are many others candidates of dark energy (quintom, quintessence, brane, modified gravity etc.).
The property of dark energy is that it is model dependent and in order to differentiate different models of dark energy,
a sensitive diagnostic tool is needed. Hubble parameter H, and deceleration parameter $q$ are very important
quantities which can describe the geometric properties of the universe. Since $\dot{a}>0$,
hence $H>0$ means that the universe expands. Also, $ddot{a}>0$, which is $q<0$ indicates the
accelerated expansion of the universe. Since, the various dark energy models give $H>0$ and $q<0$,
one needs a further evidence to differentiate general models of dark energy by investigating cosmological
observational data more accurately. For this aim, we need higher order of time derivatives of scale factor,
a geometrical tool. In the Ref. \cite{Sahni} geometrical statefinder diagnostic tool proposed, based on dimensionless
parameters $(r, s)$ which are function of scale factor and its time derivative.\\

\begin{figure}[h!]
 \begin{center}$
 \begin{array}{cccc}
\includegraphics[width=50 mm]{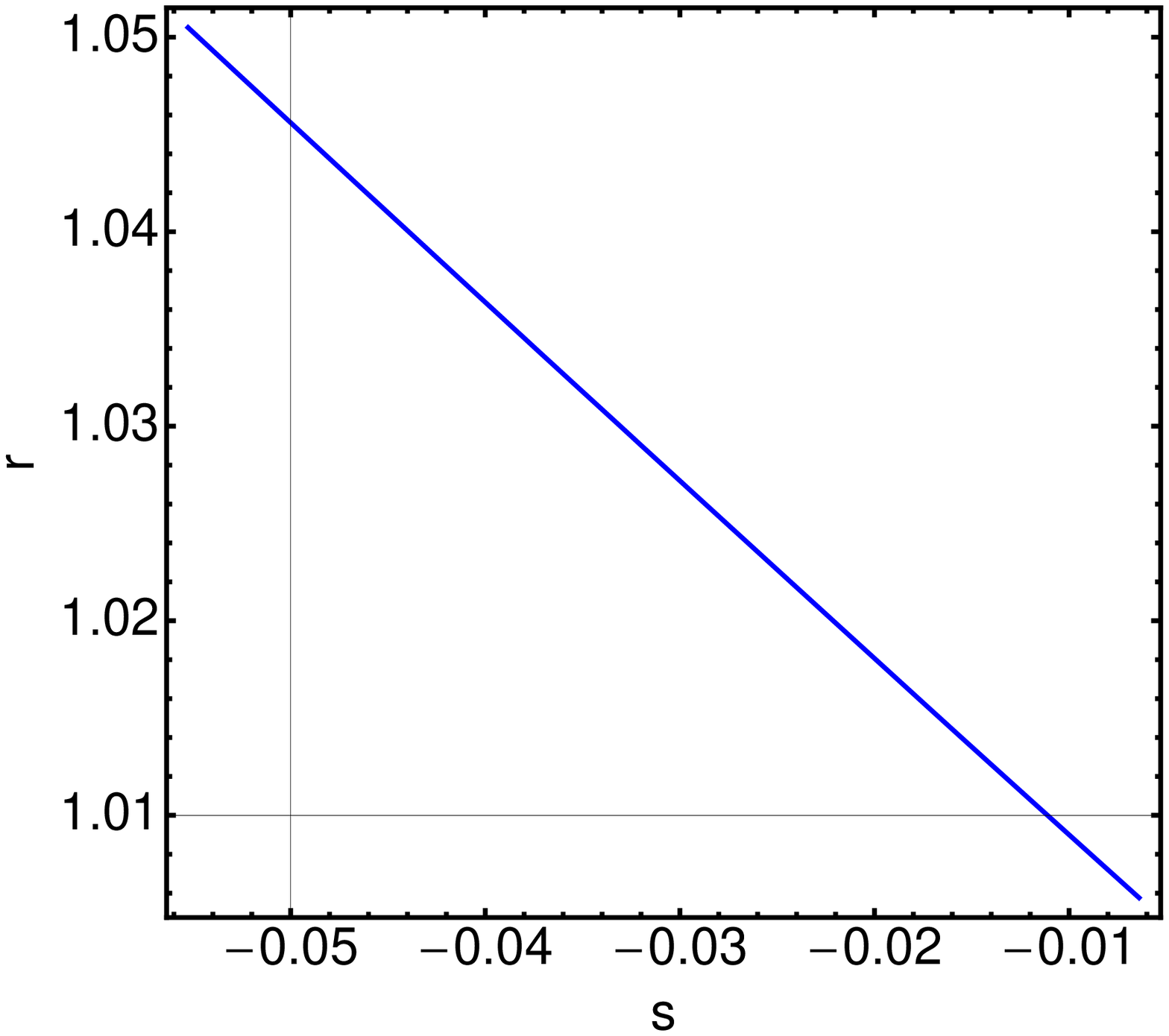} &
\includegraphics[width=50 mm]{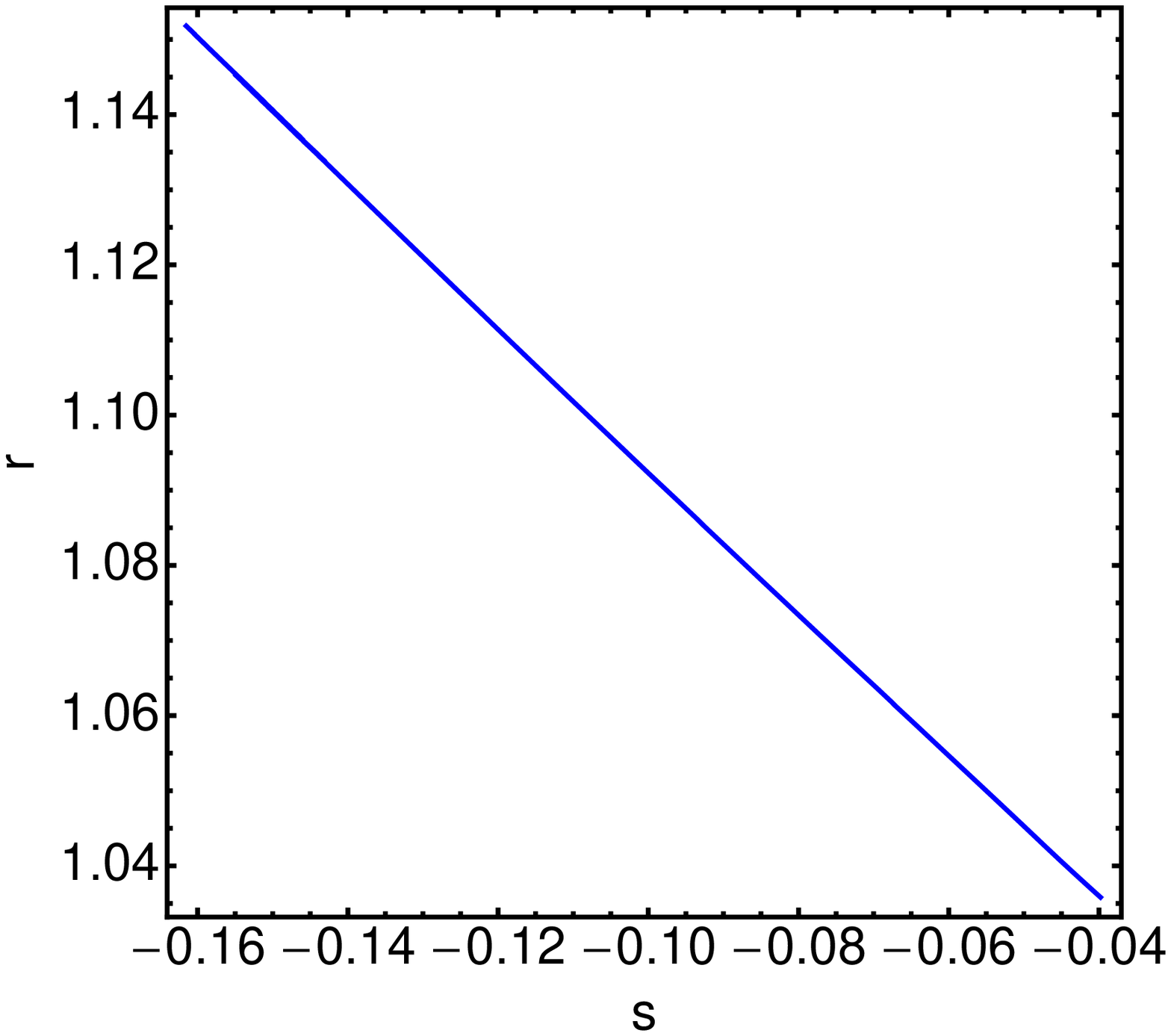}\\
\includegraphics[width=50 mm]{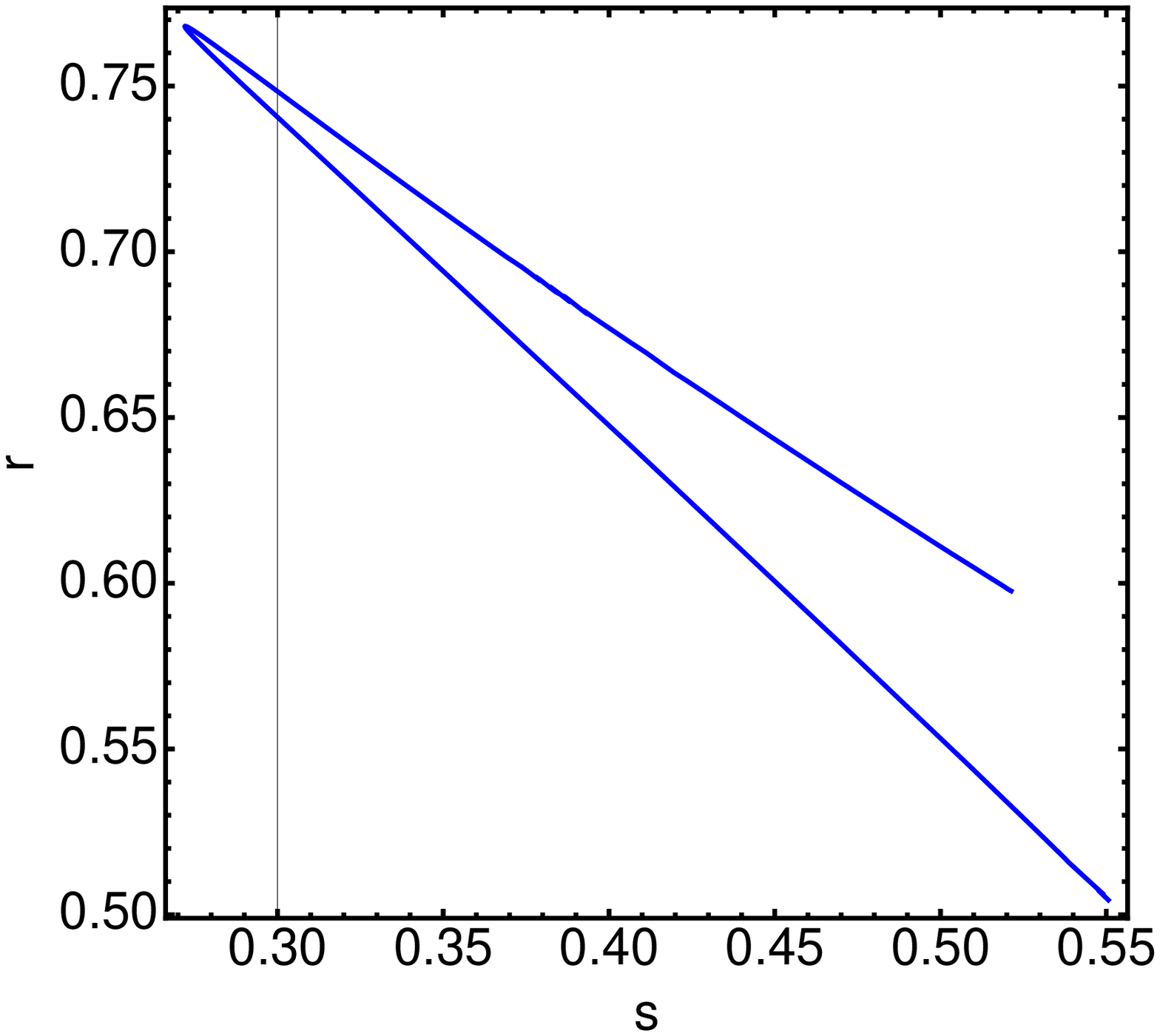} &
\includegraphics[width=50 mm]{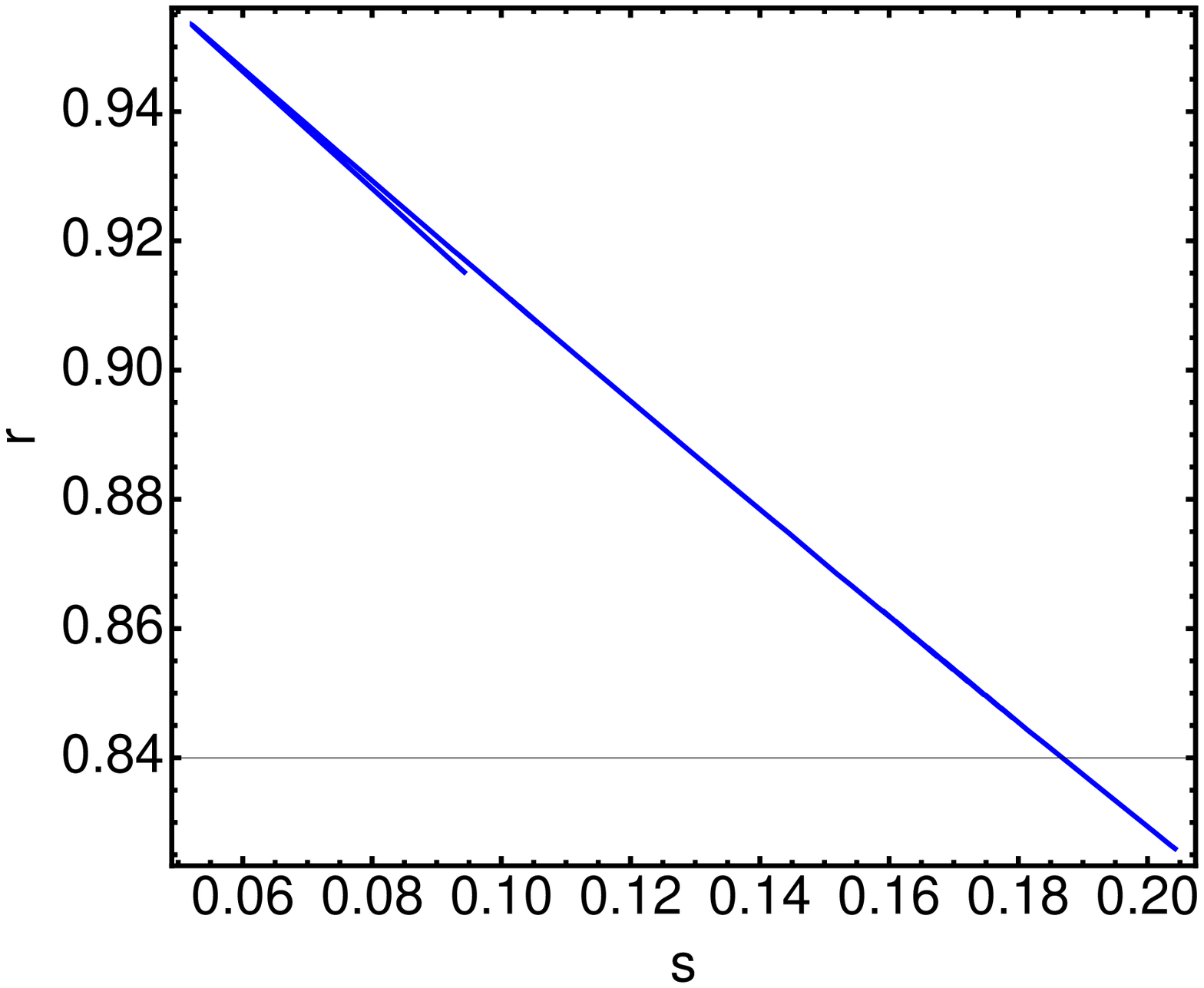}\\
 \end{array}$
 \end{center}
\caption{$r-s$. Top panels corresponds to the cases of constant $G$ and $\Lambda$ for model 1 and model 2, respectively. Bottom panels corresponds to the cases of varying $G$ and $\Lambda$ for model 1 and model 2, respectively.}
 \label{fig:12}
\end{figure}

These parameters are defined as follows:
\begin{equation}\label{24eq:statefinder}
r=\frac{1}{H^{3}}\frac{\dddot{a}}{a} ~~~~~~~~~~~~
s=\frac{r-1}{3(q-\frac{1}{2})}.
\end{equation}
Plots of Fig. \ref{fig:12} show results of our numerical analysis in both models for the cases of constant ($G$, $\Lambda$), and varying ($G$, $\Lambda$). We can see that the value of $r$ in the cases of varying $G$ and $\Lambda$ are lower than the cases of constant $G$ and $\Lambda$. However $(r, s)=(1, 0)$ verified in all cases.

\section{Perturbation analysis}
\begin{figure}[h!]
 \begin{center}$
 \begin{array}{cccc}
\includegraphics[width=60 mm]{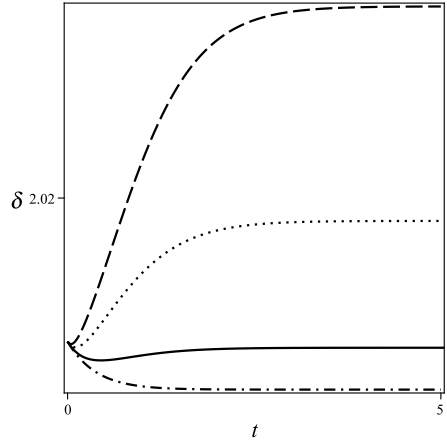}
 \end{array}$
 \end{center}
\caption{Typical behavior of $\delta$ in terms of $t$ for the extended Chaplygin gas. $m=1$ (solid), $m=2$ (dotted), $m=3$ (dashed).}
 \label{fig:13}
\end{figure}

\begin{figure}[h!]
 \begin{center}$
 \begin{array}{cccc}
\includegraphics[width=50 mm]{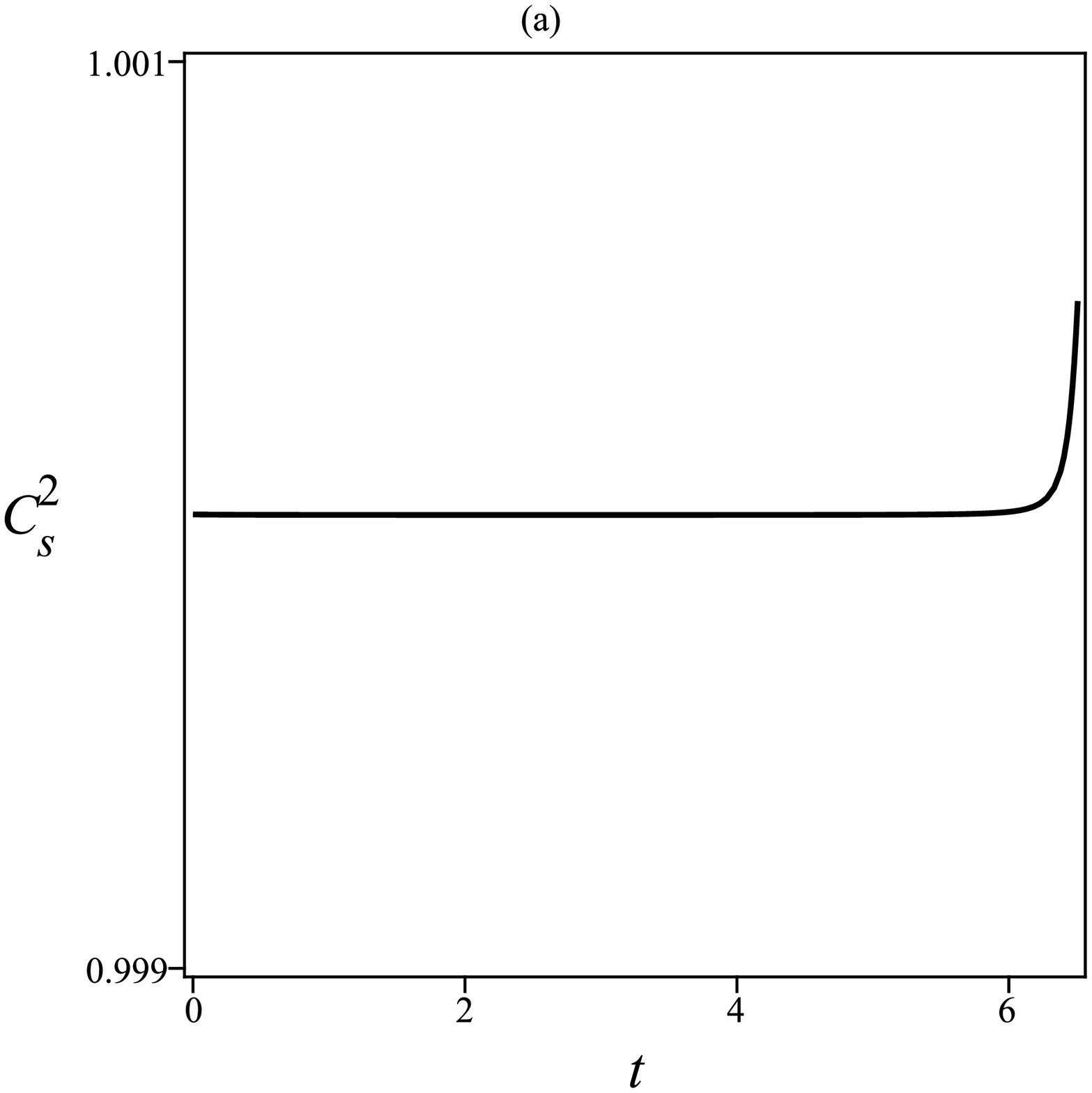}\includegraphics[width=50 mm]{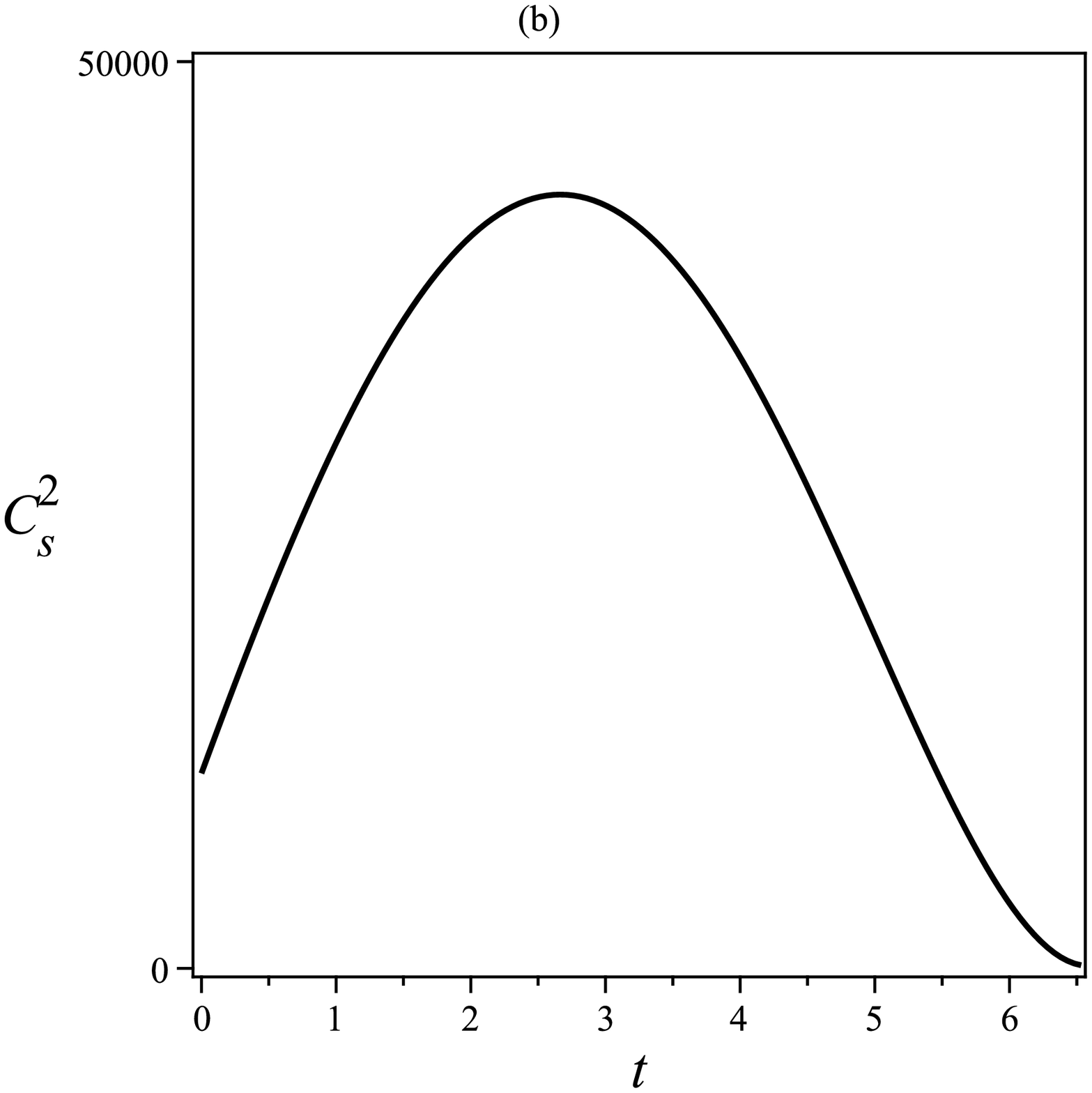}\includegraphics[width=50 mm]{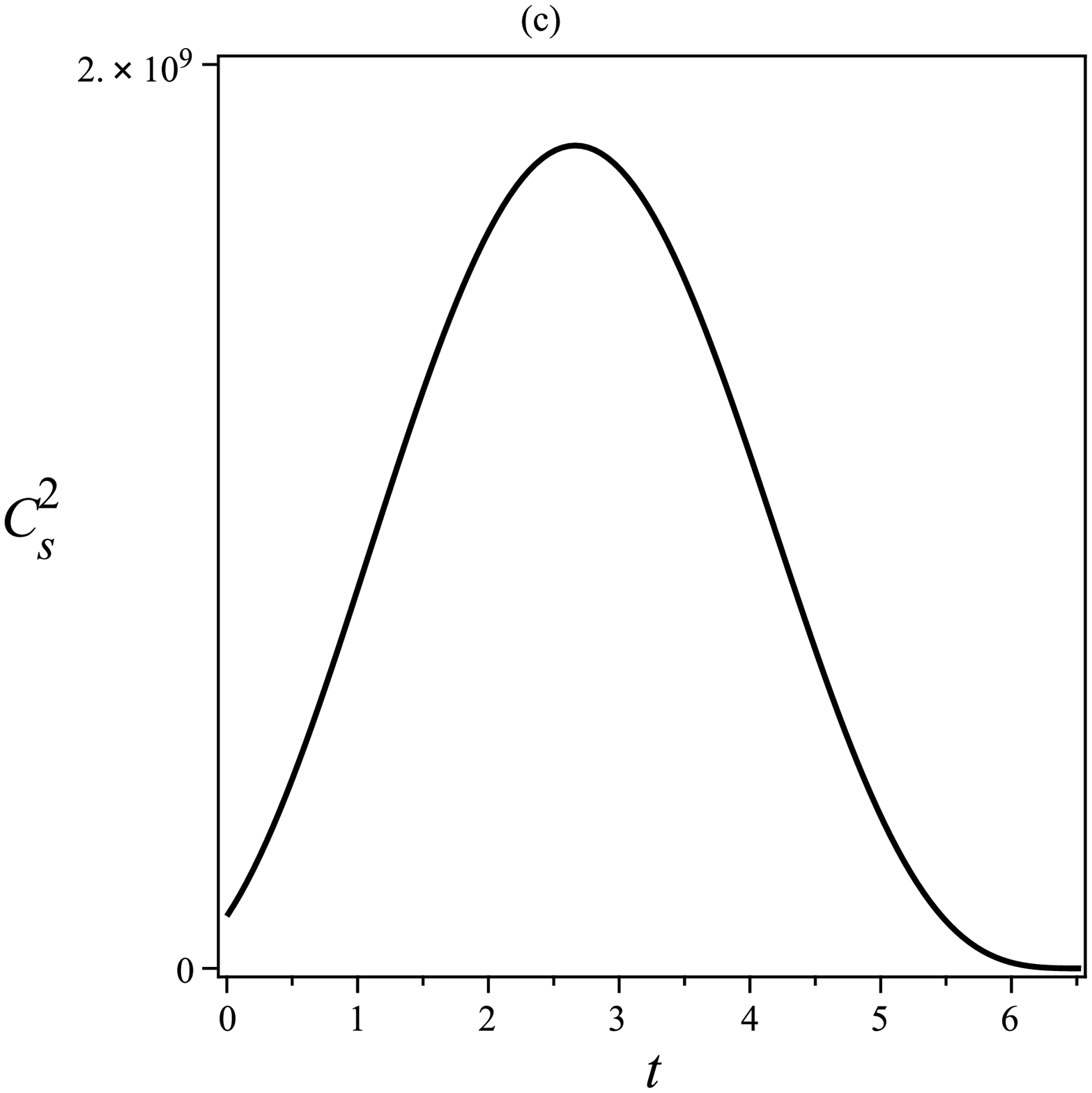}
 \end{array}$
 \end{center}
\caption{Typical behavior of squared sound speed in terms of $t$ for the extended Chaplygin gas. (a) $m=1$, (b) $m=2$, (c) $m=3$.}
 \label{fig:14}
\end{figure}

One of the best ways to investigate the stability of a model is studying cosmological perturbations \cite{MNRAS325(2001)}. The first equation is obtained using the continuity equation given in Eq. (\ref{14eq:constlambda}):
\begin{equation}\label{25eq:constlambdapert}
\delta\dot{\rho}=-3\delta H(\rho+P)-3H(1+C_{s}^{2})\delta\rho,
\end{equation}
where
\begin{equation}\label{26soundspeed}
C_{s}^{2}=\frac{\delta P}{\delta\rho}
\end{equation}
is the squared speed of the sound. On the other hand, using Eq. (4), we can obtain the following relation:
\begin{eqnarray}\label{27eq:Fridmmanvlambda2pert}
\delta\dot{H}&=&-2H\delta H+6\alpha\delta(2H\ddot{H}-\dot{H}^{2}+6\dot{H}H^{2})  \nonumber \\
&&+\frac{4\pi}{3}\left[\frac{\gamma}{8\pi}\delta(1+3\omega)\rho-G(1+C_{s}^{2})\delta\rho\right]+\frac{\gamma}{3}\delta\rho,
\end{eqnarray}
where we used Eqs. (\ref{18eq:lambda}) and (\ref{20eq:lambda2}) together with $P=\omega\rho$ and $\delta=\frac{\delta\rho}{\rho}$. We can then rewrite Eq. (\ref{13eq:Fridmmanvlambdaa}) in first order in $\delta H$, and
differentiate it with respect to $t$, obtaining:
\begin{eqnarray}\label{28eq:Fridmmanvlambdapert}
\delta\dot{H}&=&\left[3\alpha\frac{2H\ddot{H}-\dot{H}^{2}+6\dot{H}H^{2}}{H}-H+\frac{4\pi G}{3H}\rho\right]\ddot{\delta}\nonumber\\
&+&\left[3\alpha\frac{d}{dt}(2H\ddot{H}-\dot{H}^{2}+6\dot{H}H^{2})-\dot{H}+(1+C_{s}^{2}+(1+\omega)\rho)(\gamma+4\pi G)\right]\dot{\delta},
\end{eqnarray}
where we used $\delta=\frac{\delta H}{H}$ along with Eq. (\ref{25eq:constlambdapert}). We are interested in the case of the extended Chaplygin gas; so combining Eqs. (\ref{27eq:Fridmmanvlambda2pert}) and (\ref{28eq:Fridmmanvlambdapert}) gives:
\begin{equation}\label{29pertEQ}
X\ddot{\delta}+Y\dot{\delta}+Z\delta=0,
\end{equation}
where
\begin{eqnarray}\label{30coeffpertEQ}
X&\equiv&\left(5-2\frac{\dot{H}}{H^{2}}\right)\left[3\alpha\frac{2H\ddot{H}-\dot{H}^{2}+6\dot{H}H^{2}}{H}-H+\frac{4\pi G}{3H}\rho\right],\\
Y&\equiv&\left(5-2\frac{\dot{H}}{H^{2}}\right)\left\{3\alpha\frac{d}{dt}(2H\ddot{H}-\dot{H}^{2} +6\dot{H}H^{2}) \right. \nonumber \\
&& \left.-\dot{H}+\left[1+C_{s}^{2}+(1+\omega)\rho\right](\gamma+4\pi G)\right\},\\
Z&\equiv&12\dot{H}+2\frac{\ddot{H}}{H}+\frac{\gamma}{6\pi}(1+3\omega)-\frac{4\pi G}{3}(1+C_{s}^{2})+\frac{\gamma}{3}-2.
\end{eqnarray}
Then, using Eqs. (4) and (\ref{23}) we can obtain a time-dependent equation to investigate perturbation evolution. We can see from the Fig. \ref{fig:13} that perturbations will grow for $m>1$. However, after long time, they lead to a constant. Moreover, the analysis of the squared speed of the sound $C_{s}^{2}$ shows that the model is stable at all time. In the Fig. \ref{fig:14} we have plotted the behavior of the squared speed of the sound for three different values of $m$, in particular $m=1$, $m=2$ and $m=3$.\\

In the case of $m = 1$ which corresponds to MCG we have constant squared sound speed at initial time
increasing at the late time. For the case of $m = 2$, i.e. quadratic barotropic fluid,
we can see that value of squared sound speed increases dramatically at the early universe
and decreasing at later stages to a constant value. Similar behavior is observed for the case of $m = 3$.
Therefore, we can conclude that squared sound speed is positive for these models hence stable. 
\section{Conclusion}
In this work, we considered higher order $f(R)$ gravity with time-dependent $G$ and $\Lambda$. We assumed Chaplygin gas as a candidate for dark energy and considered two models to describe the evolution of our universe. The first one was the variable modified Chaplygin gas model (VMCG). The constant $B$ in ordinary MCG, is a time-dependent quantity in the context of VMCG. The second model was the extended Chaplygin gas. This is indeed extended version of MCG to recover higher order barotropic fluid EoS.\\
We assumed a specific case where $\Lambda$ being proportional to the energy density and analyzed the Hubble, the deceleration and the EoS parameters. We found that varying $G$ and $\Lambda$ fit the observational data compared to the cases of constant $G$ and $\Lambda$.\\
We also found that the extended Chaplygin gas is a more appropriate model compared to the variable modified Chaplygin gas model. We can see also acceleration to deceleration phase transition in the extended Chaplygin gas model.\\
Finally, we investigated the evolution of the density perturbations by looking at perturbed Friedmann equations.
We also confirmed stability of the extended Chaplygin gas by investigating the square of the speed of sound.
In summary, we think that the higher order corrected extended Chaplygin gas with
time-dependent $G$ and $\Lambda$ as a possible model able to describe the evolution of our universe.

\end{document}